\documentclass[11pt, hidelinks]{article} 

\usepackage{hyperref}
\hypersetup{colorlinks,linkcolor={blue},citecolor={blue},urlcolor={red}}

\usepackage{url}
\usepackage{tcolorbox}
\usepackage{fullpage}
\usepackage{algorithmic}
\usepackage{color,xcolor}
\usepackage{xspace}

\usepackage{siunitx}
\usepackage[mathscr]{eucal}
\usepackage{amsbsy}

\usepackage[linesnumbered,ruled]{algorithm2e}

\usepackage{fancyhdr}

\usepackage{microtype}

\usepackage{bbm,amsmath,amssymb,amsthm, amsfonts}

\usepackage{mathtools}

\usepackage{array,multirow}

\usepackage{enumitem}

\usepackage{comment}

\usepackage{xifthen}

\usepackage[noabbrev,nameinlink]{cleveref}
\crefname{ineq}{Inequality}{inequalities}
\crefname{thm}{Theorem}{theorems}
\creflabelformat{ineq}{#2{\upshape(#1)}#3} 

\usepackage{wrapfig}

\DontPrintSemicolon

\usepackage{basic_latex_macros}

\usepackage{flora_macros}

\theoremstyle{plain}
\newtheorem{theorem}{Theorem}

\newtheorem{corollary}{Corollary}
\newtheorem{proposition}{Proposition}

\theoremstyle{definition}

\newcommand{\figdir}{Figure/}

\newcommand{\pUpper}{\ensuremath{\priorNull^*}}
\newcommand{\LinFun}{\ensuremath{\ell}}

\newcommand{\Bone}{(I)\xspace}
\newcommand{\Btwo}{(II)\xspace}

\newcommand{\Reward}{\ensuremath{R}}

\newcommand{\condind}{\mathrel{\text{\scalebox{1.07}{$\perp\mkern-10mu\perp$}}}}

\newcommand{\riskpar}{\ensuremath{\gamma}}

\newcommand{\altweight}{\omega_1}
\newcommand{\nullweight}{\omega_0}
\newcommand{\Putil}{\ensuremath{\mathcal{V}}}
\newcommand{\PutilBar}{\underbar{\Putil}}
\newcommand{\Util}{\util}

\usepackage[giveninits=true,  
  maxnames = 2,  
  backref=true,
  backend=bibtex,
  style=alphabetic,
  sortlocale=de_DE,
  natbib=true,
  doi=false,
  isbn=false,
  url=false]{biblatex}

\begin{document}

\begin{center}

  {\bf{\LARGE Sharp Results for Hypothesis Testing with Risk-Sensitive
      Agents}} \\

  \vspace*{.2in}
  \begin{tabular}{ccc}
    Flora C. Shi$^\dagger$ & Stephen Bates$^\dagger$ & Martin
    J. Wainwright$^{\dagger, \star}$
  \end{tabular}
  
  \vspace*{0.1in}
  
  \begin{tabular}{c}
    Laboratory for Information and Decision Systems \\
    Statistics and Data Science Center \\
    EECS$^{\dagger}$ and Mathematics$^{\star}$ \\
    Massachusetts Institute of Technology
  \end{tabular}
  
  \vspace*{.2in}

  \today

  \medskip 
  \begin{abstract}
    Statistical protocols are often used for decision-making involving multiple parties, each with their own incentives, private information, and ability to influence the distributional properties of the data. We study a game-theoretic version of hypothesis testing in which a statistician, also known as a principal, interacts with strategic agents that can generate data. The statistician seeks to design a testing protocol with controlled error, while the data-generating agents, guided by their utility and prior information, choose whether or not to opt in based on expected utility maximization. This strategic behavior affects the data observed by the statistician and, consequently, the associated testing error. We analyze this problem for general concave and monotonic utility functions and prove an upper bound on the Bayes false discovery rate (FDR). Underlying this bound is a form of prior elicitation: we show how an agent's choice to opt in implies a certain upper bound on their prior null probability. Our FDR bound is unimprovable in a strong sense, achieving equality at a single point for an individual agent and at any countable number of points for a population of agents. We also demonstrate that our testing protocols exhibit a desirable maximin property when the principal's utility is considered. To illustrate the qualitative predictions of our theory, we examine the effects of risk aversion, reward stochasticity, and signal-to-noise ratio, as well as the implications for the Food and Drug Administration's testing protocols.
  \end{abstract}
\end{center}


\section{Introduction}
\label{secIntro}

Statistical protocols are fundamental to various regulatory approval processes, serving as the backbone for validating the efficacy and reliability of new products or interventions. For example, hypothesis testing is widely employed to analyze data from clinical trials during drug approval processes. Scientific journals rely on $p$-values to assess the significance of study results. Technology and financial companies use A/B testing to evaluate the performance of new features. Meanwhile, the regulatory landscape is inherently complex, characterized by the presence of multiple stakeholders, each with distinct goals and incentives.

In this environment, agents such as pharmaceutical companies, researchers, and engineers typically bear the costs of collecting evidence to support their proposals. As a consequence, agents are likely to make strategic decisions, such as determining what evidence to gather and which proposals to advance, with the aim of maximizing their returns. These returns may take the form of financial gain, career advancement, or professional prestige. On the other hand, regulators seek to design decision-making protocols that achieve controlled statistical error and high social utility. A key challenge lies in the interaction—and possible misalignment—between these goals. The statistical protocols used by regulators inform decision-making and reward distribution, thereby impacting the interests of the involved agents. In many cases, the chosen protocol can incentivize agents to act in ways that are not aligned with the regulator's goals. These challenges are often exacerbated by information asymmetry between agents and regulators.

In this paper, we study a game-theoretic formalization of statistical decision-making with strategic agents, known as \emph{principal-agent hypothesis testing}. In this framework, the statistician—also referred to as the principal or regulator—seeks to design a hypothesis testing protocol with controlled error. To achieve this, she interacts with a population of agents capable of generating data and possessing prior information (known only to the agents) about its quality. The agents' preferences are represented by a utility function. For a given testing protocol, each agent decides whether or not to opt in based on maximizing their expected utility. The statistician observes data only from agents who opt in; consequently, the agents' strategic behavior influences the distribution of the data observed by the statistician and, therefore, the associated testing error.

As a concrete illustration, consider the hypothesis testing procedures employed by the Food and Drug Administration (FDA) for approving medical treatments. In its regulatory role, the FDA acts as the statistician, or principal: it issues guidelines outlining how evidence from clinical trials will be evaluated, specifying a 
$p$-value threshold for testing. On the other side, pharmaceutical companies are the agents: they develop potential drug candidates and face the decision of whether to conduct clinical trials, the cost of which can range from millions to billions of dollars. If the FDA approves a drug based on trial evidence, the company can market the drug and generate substantial profits.

Given their internal research divisions, companies possess considerable prior information about the possible effectiveness of proposed drugs; this information is not available to the FDA, creating a form of information asymmetry. In addition to this prior information, companies make decisions based on the potential profits associated with an approved drug. The possibility of high profits provides an incentive to conduct trials even for drugs with uncertain effectiveness, thereby increasing the false positive rate for the FDA's testing protocol.  At the same time, most companies exhibit various forms of risk aversion~\citep{giambona2018theory}, including avoidance of reputational damage or legal liability, and accounting for high uncertainty or delays in future profits. Such risk-averse behavior also influences the data observed by the FDA. Overall, the FDA's testing error is determined by a delicate interaction between the incentives created by any protocol and companies' prior information and risk preferences.

\subsection{How to control testing error with strategic agents?}

Thus, we are led to the central question tackled in this paper: \emph{for a given hypothesis test proposed by the principal, how can the probability of false positives be controlled?} This question is both conceptually interesting, as it quantifies the delicate interplay between incentives and statistical performance described above, and practically important, as it provides guidance for designing statistical tests with guaranteed error control. We analyze this question in a Bayesian setting, where the null and alternative spaces are endowed with priors known only to the agents. In this context, it is most natural to control the Bayesian false discovery rate (FDR), which corresponds to the posterior probability of a false positive; see equation~\eqref{EqnBayesFDR} for the precise definition.

A notable feature of our framework, compared to past work on principal-agent testing~\cite{tetenov2016economic,bates2023incentivetheoretic,bates2022principalagent}, is its incorporation of both risk aversion and uncertainty in future rewards. More specifically, we allow for agents who make decisions based on an arbitrary concave and increasing utility function. This family includes risk-averse agents, who are influenced by the stochasticity of future rewards. By contrast, the risk-neutral agents analyzed in prior work—whose underlying utility is linear—are indifferent to such uncertainty. Within this general framework, our main result, stated as~\Cref{ThmBound}, provides an upper bound on the Bayesian FDR for any choice of the principal's hypothesis test. When specialized to risk-neutral agents (linear utility) and constant rewards, this general bound offers a tighter guarantee than prior work by a subset of the current authors~\cite{bates2023incentivetheoretic}. This improvement is quantified in~\Cref{CorLinearBound} and illustrated in~\Cref{FigRiskNeutral}.

Remarkably, the upper bound provided in this paper is \emph{unimprovable} in a rather strong sense. First, in the simpler setting where the principal interacts with a single agent, we prove that the upper bound on the Bayes FDR is achieved with equality at some point; see~\Cref{ThmBound}(b) for the precise statement. When the principal interacts with a population of agents, we establish an even stronger result—namely, there exists an agent population for which the bound can be made $\epsilon$-tight at any countable number of points. See~\Cref{ThmStaircase} and the accompanying~\Cref{FigStairCase} for an illustration of this ``staircase sharpness" guarantee. Furthermore, these guarantees have significant implications for the principal's social utility. In particular, we demonstrate how to design a test that is maximin optimal for the principal (see~\Cref{PropMaximin}).

Finally, an interesting by-product of our analysis is a connection to the notion of prior elicitation (see~\Cref{SecRelated} for related work). In particular, as an intermediate step in our proof, we show how, for a given testing protocol, an agent's decision to opt in implies a certain upper bound on their prior probability of being null. Thus, the principal can be viewed as eliciting information about the agent's prior through her design of the testing protocol and the associated payoff structure. See~\Cref{PropPriorElicit} for the details of this result.


\subsection{Related work}
\label{SecRelated}

The principal-agent form of interaction is a classical game-theoretic
model in contract
theory~\citep{laffont2001,bolton2004contract,salanie2005economics}.
It can be understood as a Stackelberg game in which the principal
moves first by declaring a protocol, and the agents produce the best
responses according to their personal utility. Our work falls within
the broader literature that uses game-theoretic approaches to
characterize how strategic behavior (including incentive misalignment)
can influence statistical procedures.  As one instance of this type of
interplay, there is a rich literature on the phenomenon of
``$p$-hacking'' within the publication process: researchers
selectively report statistically significant results, thereby
increasing the likelihood of false positives.  Classical work on this
problem
(e.g.,~\citep{sterling1959publication,tullock1959publication,leamer1974false})
studied how reported estimates reflect not only data but also
researcher preferences; various corrections for such bias have been
proposed in the statistical literature on selective inference
(e.g.,~\citep{taylor2015statistical,berk2013valid}).  Most related to
this paper is the econometrics literature, in which researchers tackle
$p$-hacking from a game-theoretic perspective, modeling incentive
structures explicitly. For instance, \citet{mccloskey2024critical}
provided critical values that remain robust even when researchers engage
in $p$-hacking. Other studies also focus on scenarios in which agents
could influence data collection or specify part of the statistical
protocol. \citet{chassang2012selective,ditillio2017} examined
randomized controlled trials where agents' hidden actions affect
outcomes.  Working within the principal-agent formalism,
\citet{spiess2018optimal} studied a problem in which the principal
delegates statistical inference tasks to the agent, and recommended
restricting the agent to fixed-bias estimators for accurate estimates.
\citet{mcclellan2022experimentation} suggested adjusting the approval
threshold sequentially to incentivize continued experimentation.

The version of principal-agent testing studied here was introduced
by~\citet{tetenov2016economic}, who considered risk-neutral agents,
bounded the type I error in terms of a cost-profit ratio, and provided a certain type of maximin guarantee.  We discuss connections
to this latter guarantee in~\Cref{SecMain:maximin_optimality}.
Working in a Bayesian setting, \citet{bates2023incentivetheoretic}
proved an upper bound on the false discovery rate for risk-neutral
agents, whereas~\citet{bates2022principalagent} demonstrated how the principal can utilize $e$-values in statistical protocols to protect
against null strategic agents. Moving from binary to multiple
hypothesis testing, \citet{viviano2024modelmultiplehypothesistesting}
also considered incentive misalignment, and characterized the optimal
critical value based on the agent's cost and the number of hypotheses.

As discussed previously, underlying our analysis is a characterization
of the principal's ability to elicit the agent's private information.  For
this reason, our work also has connections to the literature on
information elicitation mechanisms. These studies focus on designing
payment structures to ensure truthful reporting from the agent; for
example, see the paper~\cite{kong2019information} on
information-theoretic approaches to designing systems for information
elicitation.  An important class of elicitation techniques includes
proper scoring rules (e.g.,~\citep{brier1950verification,good1952rational,savage1971elicitation,gneiting2007strictly}), which
are used to elicit probabilistic estimates from risk-neutral agents.
Unlike proper scoring rules, our model does not assume that the
principal can adjust the agent's rewards and costs, nor does it
require the agent to report their private information
directly. Instead, this information is inferred through an analysis of
the agent's actions.  From another perspective, the agent's decision
to opt in can be viewed as a signal to the principal, making
research relevant to the growing literature
(e.g.,~\cite{carpenter2007regulatory,andrews2021model,banerjee2020theory,henry2019research,williams2021preregistration})
on persuasion and signaling in scientific communication.

\paragraph{Paper organization:} The remainder of this paper is
organized as follows.  We begin in~\Cref{SecSetup} with a precise
formulation of principal-agent testing.  \Cref{SecMain} is devoted to
our main results.  To clarify connections with past work, we
start in~\Cref{SecMain:neutralAgents_constReward} by discussing the
consequences of our general results for the special case of
risk-neutral agents and constant rewards.  \Cref{SecMain:General} is
devoted to the statement of our main result (\Cref{ThmBound}), which
provides a general upper bound on the Bayes FDR of the principal's test.
In~\Cref{SecMain:staircase_sharpness}, we prove the
staircase-sharpness property of this upper bound,
whereas~\Cref{SecMain:maximin_optimality} is devoted to maximin
properties.  In~\Cref{SecApplication}, we illustrate some consequences
of our theory, both for synthetic problems and for the testing
problem faced by the FDA.  We conclude with a discussion
in~\Cref{SecDiscussion}.

\section{Hypothesis Testing with Strategic Agents}
\label{SecSetup}

In this section, we provide a precise formulation of the problem under study, including the structure of the hypothesis spaces, the principal-agent interaction, and the utility functions that drive the agents' decision-making.

\subsection{Principal-agent actions and the hypothesis test game}

We consider a game-theoretic form of interaction involving two parties: the principal, who acts as a statistical regulator, and the agent. The principal's role is to decide whether to approve or deny any proposal submitted by the agent. Accordingly, we represent the principal's action space as $\{\approves,\denies\}$. The interactive process begins with the principal, who declares the form of the statistical hypothesis test she will use to make her decision. Based on shared knowledge of this test, as well as their private information and utility (discussed below), the agent decides whether to submit a proposal or opt out. Accordingly, we represent the agent's action space as 
$\{\optin,\optout\}$.

\subsubsection{Hypothesis space and priors}

To formalize the binary testing problem, we assume that there exists a
hidden random variable $\theta$, which takes values in some space $\Theta$ and represents the latent quality of any proposal.  Neither the agent
nor the principal knows the true value of $\theta$, but the agent has
knowledge of a prior distribution $\priorDist$ over $\Theta$.  The
principal has no access to $\priorDist$ and cannot ask the agent
directly about $\priorDist$. (Indeed, a strategic agent has no
incentive to respond truthfully and might choose to report incorrect
information for their own benefit.)  Given the parameter space $\Theta$,
the hypothesis test is defined by a disjoint partition of the form
\begin{align}
  \label{EqnDisjoint}
 \Theta & = \underbrace{\Theta_0}_{\scriptsize{\mbox{Null set}}}\;
 \cup \; \underbrace{\Theta_1}_{\scriptsize{\mbox{Non-null or
       alternative set}}}.
\end{align}
The set $\Theta_0$ corresponds to ineffective proposals, whereas the
alternative set $\Theta_1$ corresponds to effective proposals.  An
important quantity in our analysis is the \emph{prior null
probability} $\priorNull \defn \priorDist(\theta \in \Theta_0)$.

\subsubsection{Evidence generation and threshold choice}

The principal's goal is to design testing rules that, as much as
possible, lead to the \approves decision for effective proposals and
the \denies decision for ineffective ones.  We now specify the
form of the tests and the variables to which they apply.  Let $\{
\Pdist_\theta \mid \theta \in \Theta \}$ be a family of (conditional)
probability distributions indexed by $\theta$.  We use this family
to model both (a) the evidence $\evidence$ used by the principal for
decision-making; and (b) the random reward $\Reward$
received by an agent for approved proposals.  When an agent with
parameter $\theta$ chooses to submit a proposal, they
conduct an experiment that yields a random variable $\evidence$, drawn
according to the distribution $\Pdist_\theta$.  In typical cases, the
variable $\evidence \in [0,1]$ is a $p$-value, meaning that its
distribution under the null is uniform.  Our analysis allows for a
slight relaxation, applying to variables $\evidence$
that are \emph{super-uniform under the null}, meaning that for any
$\theta \in \Theta_0$, we have
\begin{align}
\label{EqnSuperUniform}  
  \Pdist_\theta(\evidence \leq \threshold) \leq \threshold \qquad
  \mbox{for any choice of threshold $\threshold \in [0,1]$.}
\end{align}

In order to evaluate proposals, the principal declares in
advance that she will perform a binary hypothesis test with a threshold
$\threshold \in [0,1]$---that is, she \approves the proposal if
$\evidence \leq \threshold$, and \denies it otherwise.  The principal
is not permitted to change the threshold $\threshold$ once it is
published.  Given knowledge of the threshold $\threshold$, the agent
then decides whether to spend $\cost$ dollars to run a trial (\optin)
and collect evidence $\evidence \sim \Pdist_\theta$, or to \optout.
When an agent chooses to \optin and the principal \approves the
proposal, the agent receives a randomly drawn reward $\reward
\sim \Pdist_\theta$ in dollars. Otherwise, when the principal \denies
the proposal, the agent loses $\cost$ dollars (deterministically) for
running the trials.

The sequence of interaction between the principal and the agent can be
summarized as follows:
\begin{tcolorbox}[boxrule=0.5pt,colback=white,colframe=black,sharp corners]
    {\bf Principal-agent hypothesis testing}
\begin{enumerate}[itemsep=0pt]
    \item The principal publishes her decision threshold $\threshold
      \in [0, 1]$.
    \item Based on his private information $\priorDist$, the agent
      decides whether to spend a cost of $\cost$ to \optin or \optout.
    \item If the trial is run, the agent collects evidence according
      to $\evidence \sim \Pdist_\theta$.
    \item The principal \approves the proposal if $\evidence \leq
      \threshold$ and \denies otherwise.
    \item If the principal \approves, the agent receives a reward of
      $\rfunc \sim \Pdist_\theta$.
\end{enumerate}
\end{tcolorbox}

\subsubsection{Agent rewards}
\label{SecRewards}

The formulation given here allows the reward $\reward$ to be both
random and dependent on $\theta$.  Formally, for
any value $\theta \in \Theta$, we assume that $\reward$ is drawn from
the distribution $\Pdist_\theta$.  In addition, we impose a few
regularity conditions.  First, we assume throughout that
\begin{subequations}
  \begin{align}
\label{EqnRewardCostBound}    
  \Exs_{\theta}[\reward] \geq \cost \qquad \mbox{for all $\theta \in
    \Theta$.}
  \end{align}
This \emph{non-dominating cost condition} is necessary for the principal
to observe an opt-in agent.  We also require the random reward to be
\emph{stochastically monotone} in the following sense: for each pair
$\theta_0 \in \Theta_0$ and $\theta_1 \in \Theta_1$, we have
\begin{align}
\label{EqnStochMonotone}
\Pdist_{\theta_0}(\rfunc \geq r) & \leq \Pdist_{\theta_1}( \rfunc \geq
r) \qquad \mbox{for all $r \in \real$.}
\end{align}
This condition ensures that the rewards for non-null $\theta \in
\Theta_1$ are, in a probabilistic sense, at least as good as those
associated with null $\theta \in \Theta_0$.

Finally, one would expect that the sources of reward stochasticity (e.g.,
market conditions, etc.) are (conditionally) independent of the
randomness in the evidence variable $\evidence$.  Formally, we require
that the evidence $\evidence$ and reward $\reward$ are conditionally
independent given $\theta$---that is
\begin{align}
  \label{EqnDecReward}
  X \condind \rfunc \mid \theta.
\end{align}
\end{subequations}
Condition~\eqref{EqnDecReward} implies that for any threshold $\tau
\in [0,1]$, the principal's decision $\Ind(X \leq \threshold)$ is also
conditionally independent of the reward $\rfunc$.


\subsection{Utility-maximizing agents}

Having set up the hypothesis test, rewards, and the form of
principal-agent interaction, the final ingredient is the specific
mechanism by which the agent makes decisions.  He does so to
serve his own interests, particularly by maximizing his expected
utility.

More precisely, suppose that each agent has some initial wealth
$\Wealth_0 > \cost$, and let $\wealth \mapsto \Util(\wealth)$ be a
concave and non-decreasing utility function.  It measures the value
that the agent ascribes to a particular wealth level $\wealth$.  For
instance, a risk-neutral agent is defined by the \emph{linear utility
function} $\Util(\wealth) = \wealth$.  A more general family of
utility functions consists of those with constant relative risk
aversion~\citep{arrow1965riskaversion,PrattJohn1964}, given by
\begin{align}
\label{eq:risk_averse_utility}
\util_\riskpar(\wealth) & \defn \frac{\wealth^{1 - \riskpar}}{1 -
  \riskpar} \qquad \mbox{for a risk parameter $\riskpar < 1$.}
\end{align}
By convention, we set $\util_1(\wealth) = \log(\wealth)$ for $\riskpar
= 1$; this logarithmic utility underlies betting strategies that seek
to maximize the growth rate under repeated trials (i.e., the Kelly
criterion).  Otherwise, note that the risk-neutral choice is the
special case with $\riskpar = 0$, whereas this specification captures
risk-seeking behavior for $\riskpar < 0$ and risk-aversion for
$\riskpar \in (0, 1]$. There is empirical
  evidence~\citep{giambona2018theory} showing that most companies
  exhibit some degree of risk-aversion, meaning that choices $\riskpar
  \in (0,1]$ are the most practically relevant within the
    family~\eqref{eq:risk_averse_utility}.

Given an arbitrary concave and non-decreasing utility function
$\util$, the agent chooses the action in the set $\{\optin, \optout
\}$ that maximizes expected utility.  Based on the previously specified
principal-agent interaction, we have the following three
possibilities:
\begin{itemize}
\item If the agent chooses to \optin and the principal \approves the
  proposal, the agent gains the random $\rfunc \sim \Pdist_\theta$
  dollars and has a final wealth of ${\Wealth_0 + \rfunc - \cost}$.
\item If the agent chooses to \optin and the principal \denies the
  proposal, the agent loses $\cost$ dollars and finishes with
  wealth ${\Wealth_0 - \cost}$.
\item Otherwise, if the agent chooses to \optout, his final wealth remains
  at $\Wealth_0$
\end{itemize}
These outcomes define the expected utility for both actions, and we
assume that the agents are \emph{utility-maximizing}, meaning that
they choose to \optin if and only if
\begin{align}
\label{EqnUtilMaxAgent}  
  \underbrace{\E \Big[ \util \big(\WealthAfter \big)
      \Big]}_{\mbox{\small{Expected utility of \optin}}} \quad \geq
  \quad \underbrace{\util(\Wealth_0)}_{\mbox{\small{Expected utility
        of \optout}}},
\end{align}
where $\Ind(\evidence \leq \threshold)$ is a binary indicator function
for the event $\evidence \leq \threshold$.  To be clear, the left side
of this inequality~\eqref{EqnUtilMaxAgent} involves a three-part
expectation: conditioned on the unknown value of $\theta$, we take an
expectation over the evidence $\evidence$ and reward $\reward$ drawn
from $\Pdist_\theta$; and second, we take an expectation over $\theta$
distributed according to the prior $\priorDist$ of the agent.  Thus,
as shown by our analysis, the agent's decision to \optin implicitly imposes a constraint on the prior distribution, which can be inferred by the principal.


\section{Main Results}
\label{SecMain}

We now present our main results and
discuss their consequences.  Recall that the principal moves
first by declaring a threshold $\threshold$ to be used in the
hypothesis test.  We provide guarantees on the posterior null
probability, also referred to as the Bayes false discovery rate (FDR),
associated with any such test:
\begin{align}
  \label{EqnBayesFDR}
  \mbox{Bayes False Disovery Rate (FDR):} & \qquad \posNull.
\end{align}
We have followed the terminology of \citet{efron2012large} by using
Bayes FDR.  This term is appropriate because the quantity $\posNull$
is closely related to the usual definition of false discovery, which is
the expected ratio of false positives to the total number of positives
when testing multiple hypotheses. The key difference is that our setup also
includes a Bayesian prior over the parameter space $\Theta$.

The remainder of this section is organized as follows.
In~\Cref{SecMain:neutralAgents_constReward}, we begin by presenting an upper
bound on the Bayes FDR in a simplified setting to provide
intuition for our general result and establish connections to past work.
Building on these insights, we extend our analysis
in~\Cref{SecMain:General} to derive a bound on the Bayes FDR in a more general setting
(see~\Cref{ThmBound}). This extension encompasses the fully general
framework described in~\Cref{SecSetup}, including stochastic rewards
and general utility. In~\Cref{SecMain:staircase_sharpness}, we
establish a sharpness guarantee for our upper bound,
whereas~\Cref{SecMain:maximin_optimality} provides maximin guarantees
for the principal.


\subsection{Risk-Neutral Agents and Constant Reward}
\label{SecMain:neutralAgents_constReward}

In this section, we consider a
risk-neutral agent who makes decisions using the \emph{linear
utility} function $\util(\wealth) = \wealth$, and \emph{constant
deterministic rewards}, meaning that there is a fixed scalar $\constR$ such that
$\rfunc \stackrel{(a.s.)}{=} \constR$ for all $\theta \in \Theta$.
Moreover, we consider a simple-versus-simple hypothesis test, meaning
that both the null set $\Theta_0 \equiv \{\theta_0\}$ and the non-null set
$\Theta_1 \equiv \{\theta_1 \}$ are singletons.  This scenario allows
us to demonstrate how the Bayes FDR depends on key factors such as the
cost $\cost$, the reward $\constR$, the threshold $\threshold$, and
the power of the hypothesis test.  It also facilitates concrete comparisons with past work~\cite{bates2023incentivetheoretic}.

\subsubsection{A bound on the Bayes FDR}

In this simple setting, our bound depends on four quantities: the cost
$\cost$ paid by the agent for choosing to \optin; the reward $\constR
> \cost$ received when the principal \approves; and the two functions
\begin{align}
\label{EqnDefnSimplePower}  
\beta_0(\threshold) \defn \Pdist_{\theta_0}(X \leq \threshold) \quad
\mbox{and} \quad \beta_1(\threshold) \defn \Pdist_{\theta_1}(X \leq
\threshold).
\end{align}
When $\evidence$ is actually uniform under the null, we have
$\beta_0(\threshold) = \threshold$, but our analysis only requires the
super-uniformity condition~\eqref{EqnSuperUniform}, equivalently
stated as $\beta_0(\threshold) \leq \threshold$.  Note that
$\threshold \mapsto \beta_{1}(\threshold)$ is the \emph{power
function} of the test, and we assume the test has \emph{non-trivial
power}, which means
\begin{align*}
\beta_1(\threshold) > \beta_0(\threshold) \qquad \mbox{for all
  $\threshold \in [0,1]$.}
\end{align*}

\begin{corollary}[Risk-neutral agent with constant reward]
  \label{CorLinearBound}
Under the previous conditions, the Bayes FDR associated with
risk-neutral agents who \optin for a given threshold $\threshold \in
\big [0, \: 1]$ is at most
\begin{equation}
  \label[ineq]{EqnLinearBound}
  \posNull \; \stackrel{(I)}{\leq} \; \frac{\beta_0(\threshold)
    \constR}{\cost} \; \frac{\beta_1(\threshold) \constR -
    \cost}{(\beta_1(\threshold) - \beta_0(\threshold)) \constR} \;
  \stackrel{(II)}{\leq} \; \frac{\threshold \; \constR}{\cost} \;
  \Biggr \{ \frac{\constR - \cost}{(1 - \threshold) \constR} \Biggr
  \}.
\end{equation} 
\end{corollary} 
\noindent We prove this claim using a more general result to be stated
in the sequel; see the discussion following~\Cref{ThmBound}.\\

Equation~\eqref{EqnLinearBound} provides two upper
bounds on the Bayes FDR.  Inequality (I) is a sharper result, but
computing it requires knowledge of the quantities
$\beta_0(\threshold)$ and $\beta_1(\threshold)$.  Depending on the
application, these quantities may or may not be known to the
principal. Inequality (II) is a weaker result that is established
using the facts that $\beta_0(\threshold) \leq \threshold$ by
assumption, and the upper bound $\beta_1(\threshold) \leq 1$ by
definition~\eqref{EqnDefnSimplePower}.  The advantage of the bound
(II) is that it can be calculated without knowledge of
$\beta_0(\threshold)$ and $\beta_1(\threshold)$.

When $\beta_0(\threshold) = \threshold$, inequalities (I) and (II)
provide non-trivial information---that is, an upper bound on the
posterior null less than $1$---only when $\threshold <
\cost/\constR$. As noted in prior work by~\citet{tetenov2016economic},
the fraction $\cost/\constR \in (0, 1]$ corresponds to the threshold
  at which profit-maximizing agent who knows with certainty that he is
  null would still \optin. As such, there exist (worst-case)
  testing problems for which the Bayes FDR is equal to $1$ when
  $\threshold = \cost/\constR$, so that non-trivial bounds are
  impossible above this threshold.

In past work on risk-neutral agents,
\citet{bates2023incentivetheoretic} proved that the Bayes FDR is upper
bounded by $\threshold \frac{\constR}{\cost}$.  The
bound~\eqref{EqnLinearBound} provides a refinement of
this claim: in particular, for any $\tau \in [0,
  \frac{\cost}{\constR}]$, we have
  \begin{align}
    \label{EqnChain}
  \posNull \; \stackrel{(a)}{\leq} \; \threshold \frac{\constR}{\cost}
  \Big(\frac{\constR - \cost}{\constR - \threshold \constR}\Big) \;
  \stackrel{(b)}{\leq} \; \threshold \frac{\constR}{\cost},
  \end{align}
where inequality (a) is bound (II) from
equation~\eqref{EqnLinearBound}; and inequality (b) holds since
$\frac{\constR - \cost}{\constR - \threshold \constR} \leq 1$ whenever
$\threshold \leq \frac{\cost}{\constR}$.  In fact, the
bound~\eqref{EqnLinearBound}\Bone is unimprovable in a very strong
sense; see part (b) of~\Cref{ThmBound} for a sharpness guarantee for a
general utility function.


\subsubsection{Connection to prior elicitation}

Underlying our bounds on the Bayes FDR---both in~\Cref{CorLinearBound}
and the more general~\Cref{ThmBound} in the sequel---is an interesting
connection to prior elicitation.  Here we provide an informal
description for the case of linear utility; see~\Cref{PropPriorElicit}
to follow for a precise statement in the general setting.

Consider an agent who chooses to \optin for a trial with a threshold
$\threshold \in (0, \cost/\constR]$.  This agent has their private
  prior $\priorDist$ over the parameter space $\Theta = \Theta_0 \cup
  \Theta_1$; of particular interest is the prior null probability
  $\priorNull = \priorDist(\theta \in \Theta_0)$ associated with this
  distribution.  Intuitively, the fact that this agent is
  utility-maximizing~\eqref{EqnUtilMaxAgent} and has chosen to \optin
  implies that their prior null probability cannot be too large.  In
  particular, our analysis shows that for a risk-neutral agent with
  constant reward, any agent who decides to \optin must have the prior null probability upper
  bounded
\begin{align}
\label{eq:linear_bound_on_prior_null}
\priorNull & \stackrel{\Bone}{\leq} \frac{\altAppSimple \constR - \cost}{
  \altAppSimple \constR - \nullAppSimple \constR} \; \stackrel{\Btwo}{\leq} \;
\frac{\constR - \cost}{\constR - \threshold \constR} \; = \; \frac{1 -
  \frac{\cost}{\constR}}{1 - \threshold}.
\end{align}
The bound~\Btwo is obtained from the first inequality via the upper
bounds $\beta_0(\threshold) \leq \threshold$ and $\beta_1(\threshold)
\leq 1$, and certain monotonicity properties.  See the proof
of~\Cref{PropPriorElicit} in~\Cref{AppProofPropPriorElicit} for
details.

The bounds~\eqref{eq:linear_bound_on_prior_null} reveal two phenomena
that are intuitively reasonable.  First, as the reward $\constR$
increases (with other problem parameters held fixed), these upper
bounds also increase.  Larger potential rewards for the \approves decision
mean that an agent needs less \emph{a priori} confidence in the
quality of their proposal to \optin, suggesting that their prior
null $\priorNull$ could potentially be larger.  Second, for hypothesis
tests with larger power (i.e., larger values of $\beta_1(\threshold)$),
the upper bound~\Bone increases.  Larger power implies that the agent
has more certainty that any $\theta \in \Theta_1$ they submit
will lead to the \approves decision, and hence a higher expected
profit.


\subsubsection{Illustration with Gaussian mean testing}
\label{SecGaussMean}

In this section, we examine Gaussian mean testing as an example to
compare various bounds on the Bayes FDR.

\begin{description}
\item[Gaussian mean testing:] We consider the binary hypothesis
  testing problem defined by the parameter space $\Theta = \{0,
  \theta_1\}$ with null set $\Theta_0 = \{0\}$ and non-null set
  $\Theta_1 = \{\theta_1\}$ where $\theta_1 > 0$, and the test
  statistic $Z \sim \mathcal{N}(\theta, 1)$.  We convert this test
  statistic to a $p$-value by setting $X \defn 1 - \Phi(Z)$, where
  $\Phi$ is the cumulative distribution function (CDF) of the standard
  normal.  With this choice, we have $\beta_0(\threshold) =
  \threshold$ by construction, along with the power function
\begin{align*}
\altAppSimple & = 1 - \Phi(\Phi^{-1}(1-\threshold) - \theta_1)
\end{align*}
for the alternative, where $\Phi^{-1}$ denotes the inverse CDF.
\end{description}

\begin{figure}[ht]
  \begin{center}
    \begin{tabular}{ccc}
      \widgraph{0.45\textwidth}{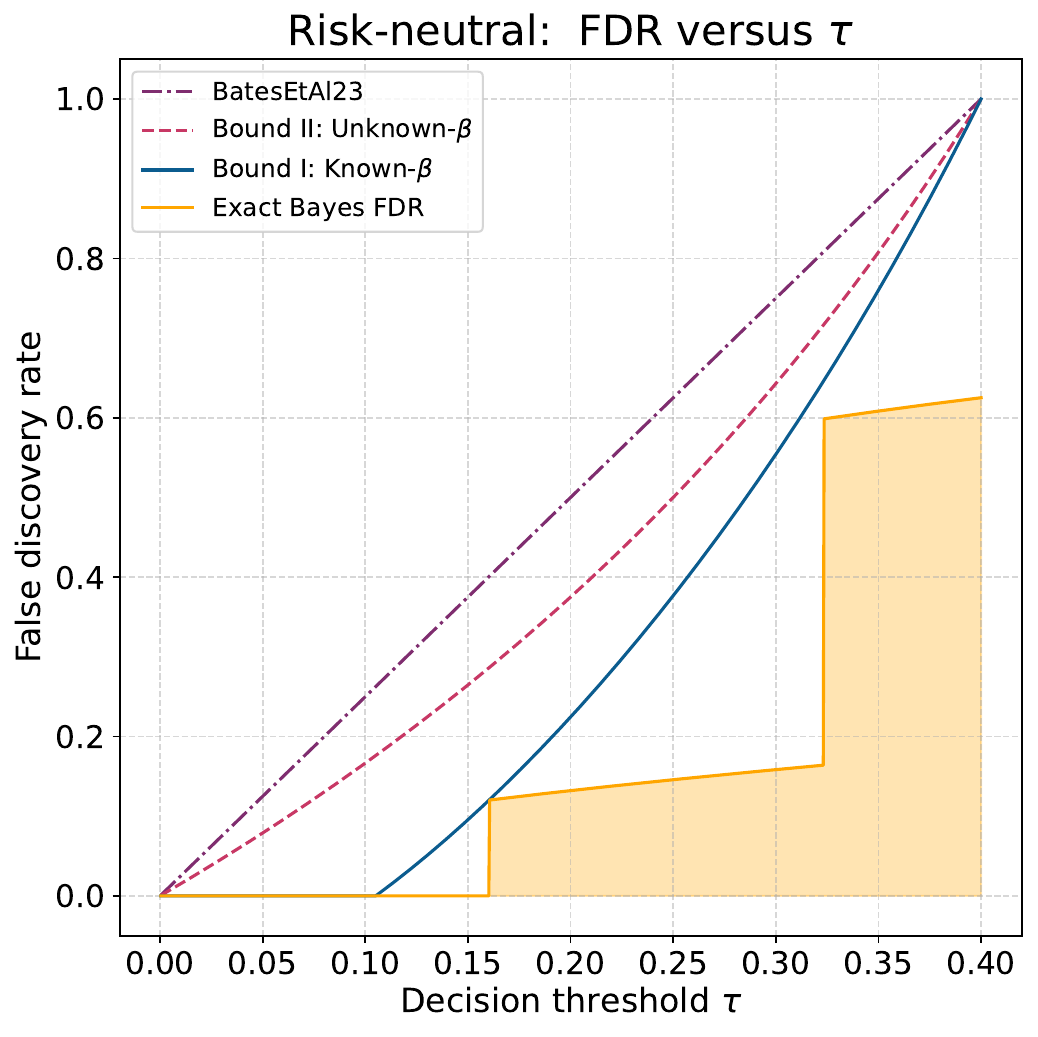}     
      &&       
      \widgraph{0.45\textwidth}{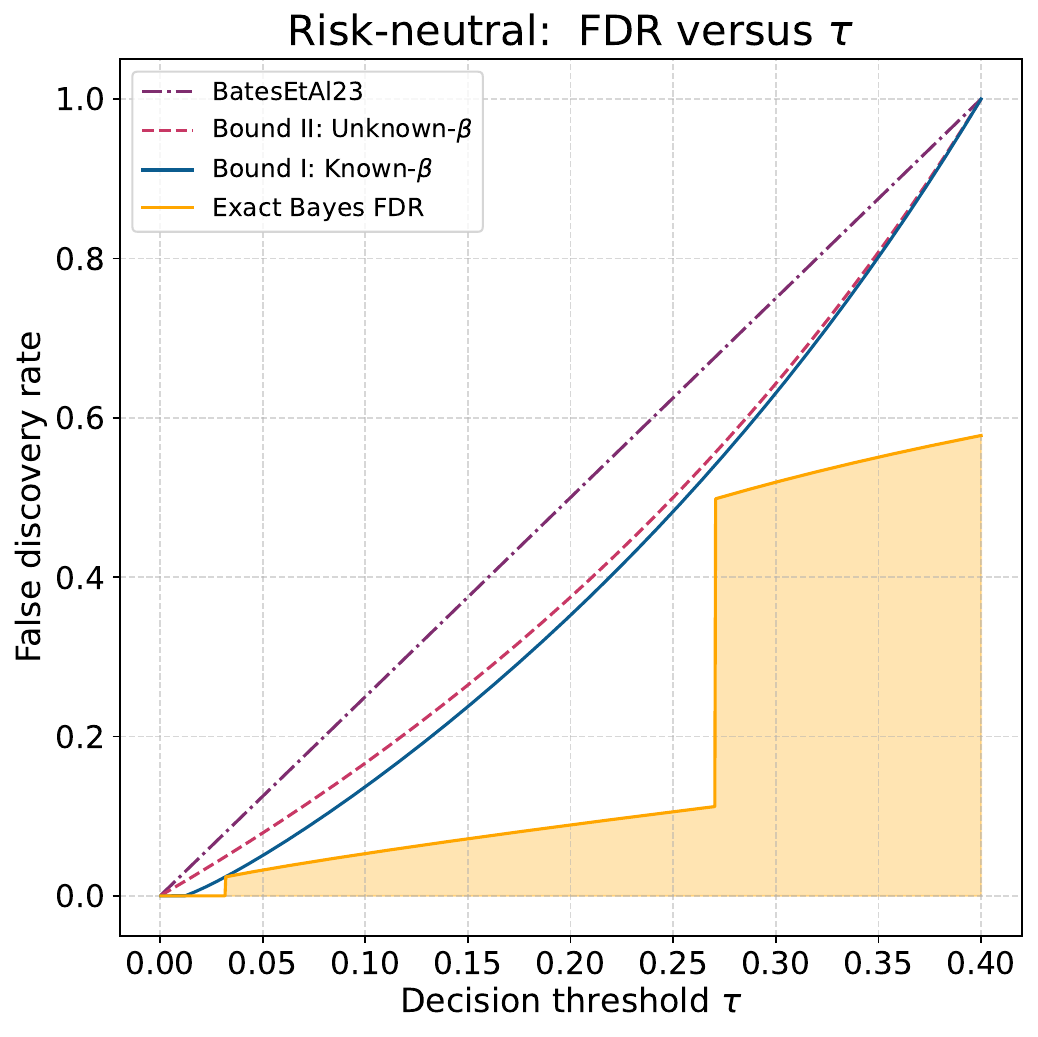} \\
      (a) Mean shift $\theta_1 = 1$ && (b) Mean shift $\theta_1 = 2$
    \end{tabular}
    \caption{Plots of the exact Bayes FDR; bounds (I) and (II) from
      equation~\eqref{EqnLinearBound}; and
      the~\citet{bates2023incentivetheoretic} bound as a function of
      the threshold choice $\threshold \in [0, \cost/\constR]$ for
      cost $\cost = 10$ and constant reward $\constR = 25$ with linear
      utility.  (a) Gaussian mean shift with $\theta_1 = 1$.  (b)
      Gaussian mean shift with $\theta_1 = 2$.}
    \label{FigRiskNeutral}
  \end{center}
\end{figure}

Given this class of hypothesis tests, we consider agents that operate
with the linear utility function, cost $\cost = 10$ and constant
reward $\constR = 25$.  For a given choice of $\theta_1$, we compute
the bounds (I) and (II) from
equation~\eqref{EqnLinearBound};
the~\citet{bates2023incentivetheoretic} bound; and the exact Bayes FDR
as a function of the threshold $\threshold$.  We plot the results for
different choices of the alternative mean $\theta_1$ in the Gaussian
testing problem: $\theta_1 = 1$ and $\theta_1 = 2$ in panels (a) and
(b), respectively, of~\Cref{FigRiskNeutral}.  For computing the exact
Bayes FDR, we consider a mixture ensemble of agent types, in which
10\% of the agents are ``good'' with $\priorNull^g = 0.3$, and the
remaining $90\%$ are ``bad'' with $\priorNull^b = 0.8$.

Focusing first on the exact Bayes FDR (yellow line with shading
underneath for emphasis) in panel (a), note that it exhibits two
step-like transitions.  The first occurs at the threshold $\threshold
\approx 0.16$; below this threshold, no agents \optin so that the FDR
is zero by definition, whereas above this threshold, agents start to
\optin.  At and above this threshold, the ``good'' agents in our
mixture ensemble \optin, and as $\threshold$ increases above this
transition, the exact Bayes FDR also increases, since the same
population of agents are provided with a progressively less stringent
hypothesis test.  The second step transition occurs at $\threshold
\approx 0.32$, at which point the ``bad'' agents also choose to \optin.  Turning to panel (b), observe that it also exhibits the same
two-step phenomenon; the difference here is that the transitions occur
for smaller values of the threshold, since the larger mean in the
alternative ($\theta_1 = 2$ in panel (b) versus $\theta_1 = 1$ in
panel (a)) means that the power function $\beta_1(\threshold)$ grows
more quickly, which leads to greater expected utility (and hence
greater incentive) for agents to \optin.

Now let us discuss the three upper bounds on the exact Bayes FDR shown
in panel (a).  By definition, the Bates et al. upper bound is linear
in the threshold $\threshold$ with slope $\cost/\constR$. Both bounds (I) and (II) from
equation~\eqref{EqnLinearBound} are non-linear and monotonic functions
of $\threshold$. Notice that bound (I) is zero near $\threshold \approx 0.1$, which represents the minimum threshold required to ensure an opt-in agent. Below this threshold, even an ideal agent with $\priorNull = 0$ will not \optin, causing the bound to evaluate to negative values. Moreover, bound (I) equals the exact Bayes FDA at the first transition, highlighting the sharpness of our bound. Consistent with our earlier analysis~\eqref{EqnChain}, both bounds
improve upon the original Bates et al. result.  The bounds in panel
(b) are qualitatively similar in nature; the main difference to note
here is that the gap between Bounds (I) and (II)---for which the
functions $\{\beta_j \}_{j=0}^1$ are known and unknown,
respectively---is smaller than in panel (a).  This difference can be
predicted by returning to the statement of~\Cref{CorLinearBound},
where we see that bound (II) is obtained by replacing the unknown
$\beta_1(\threshold)$ with the upper bound $1$. Panel (b) corresponds
to a higher signal-to-noise ratio, or SNR for short, since the size of
mean shift $\theta_1$ is doubled in moving from panel (a) and (b), so
that this approximation is more accurate for the high SNR problem. 


\subsection{General Agents and Stochastic Reward}
\label{SecMain:General}

In this section, we extend the results developed above to the more
general setting described in~\Cref{SecSetup}, allowing for a concave
and non-decreasing utility function $\wealth \mapsto \util(\wealth)$,
stochastic rewards $\reward \sim \Pdist_\theta$, and composite
hypothesis testing (so that the sets $\Theta_0$ and $\Theta_1$ need
not be singletons).  Recall from~\Cref{SecRewards} that the rewards
are assumed to satisfy three conditions. The non-dominating cost
condition~\eqref{EqnRewardCostBound} is necessary for the principal to
observe some \optin agent; the stochastic monotonicity
condition~\eqref{EqnStochMonotone} ensures that non-null hypotheses
are more rewarding than null; the conditional independence
condition~\eqref{EqnDecReward} guarantees that the evidence $\evidence$ and reward
$\reward$ are conditionally independent given $\theta$.

There are two key differences between this more general setting
and the simpler setting from the previous section.  Whereas linear
utility allows for a straightforward comparison of costs and rewards
on the wealth scale, risk-averse agents consider additional factors
beyond wealth when deciding whether to \optin. Consequently, the
agent's cost and reward must be evaluated on the utility scale. More
risk-averse agents derive less utility from monetary rewards and
therefore are less likely to \optin. Second, since the reward depends
on $\theta$, the reward when the principal \approves a non-null
proposal may differ significantly from that when the principal
\approves a null proposal. If the reward for a non-null proposal is
substantially higher, agents with greater confidence in their
proposals are more likely to \optin.

\subsubsection{Bounding the Bayes FDR}
\label{SecMain:KnownBetaGeneral}

From our discussion of the simple-versus-simple hypothesis test,
recall the two quantities $\beta_{j}(\threshold)$ for $j \in \{0,1\}$,
as defined in equation~\eqref{EqnDefnSimplePower}. Moving to the
composite-versus-composite hypothesis test, our bound depends on the
related functions
\begin{subequations}
\begin{align}
\label{EqnDefnCompositePower}
\nullApp \defn \Exs[\Ind(\evidence \leq \threshold) \mid \theta \in
  \Theta_0] \quad \mbox{and} \quad \altApp \defn \Exs[\Ind(\evidence
  \leq \threshold) \mid \theta \in \Theta_1].
\end{align}
Note for future reference that the super-uniformity
condition~\eqref{EqnSuperUniform} implies that $\nullApp \leq
\threshold$.

Observe that $\nullApp$ is a weighted average of
$\beta_0(\threshold)$, with the weights defined by the conditional
distribution of $\theta$ within the null set $\Theta_0$, denoted by
$\priorDist_0$.  On the other hand, the quantity $\altApp$ represents
the average power of the test: it is a weighted average of
$\beta_1(\threshold)$ with weights given by the conditional
distribution $\priorDist_1$ of $\theta$ over the alternative set
$\Theta_1$.  For future reference, with these definitions,
the prior distribution $\priorDist$ can be decomposed as
the two-component mixture
\begin{align}
\label{EqnMixturePrior}
\priorDist & = \priorNull \priorDist_0 + (1 - \priorNull) \priorDist_1
\end{align}
Finally, consistent with the single-versus-single case, we assume that
there is non-trivial average power---viz.
\begin{align}
\label{EqnNonTrivialCompPower}
\altApp > \nullApp \qquad \mbox{for all
  $\threshold \in [0,1]$.}
\end{align}
\end{subequations}

For a general utility function, our results depend on three forms of
difference in expected utility.  The first two correspond to
differences between the agent's expected utility, comparing the
\approves to the \denies decision.  More precisely, we define
\begin{align*}
\myemph{Utility differences:} & \qquad \UtilGain{j} \defn
\E[\util(\WealthGain) \mid \theta \in \Theta_j] - \util(\WealthLose)
\qquad \mbox{for $j \in \{0, 1 \}$.}
  \end{align*}
Note that $\util(\Wealth_0 - \cost)$ is the utility of an agent who
decides to \optin but receives a \denies decision, whereas the
quantity $\E[\util(\WealthGain) \mid \theta \in \Theta_j]$ is the
agent's expected utility for an \approves decision, taken over
$\Theta_j$ for $j \in \{0,1\}$.  Finally, the loss incurred by the
agent when the principal \denies, for $\theta$ in either the null or
the alternative, is given by
\begin{align*}
\myemph{Utility loss for \denies:} & \qquad \utilLoss \defn
\util(\Wealth_0) - \util(\WealthLose).
\end{align*}

\vspace*{0.05in}

The following guarantee applies to a principal-agent testing problem
with rewards satisfying conditions~\eqref{EqnRewardCostBound}
through~\eqref{EqnDecReward}, as well as the power
condition~\eqref{EqnNonTrivialCompPower}.

\begin{theorem}[Sharp control on posterior null]
  \label{ThmBound}
Given a concave and non-decreasing utility function $\wealth \mapsto
\util(\wealth)$, consider some threshold $\threshold \in (0, 1)$ for
which $\bar \beta_0(\threshold) \in \big(0, \; \utilLoss/\nullUtilGain
\big)$.
\begin{enumerate}[label=(\alph*), ref=\thetheorem(\alph*)]
\item \label[thm]{th:bound_on_pos_null} If a utility-maximizing agent
  decides to \optin at level $\threshold$, then the posterior null
  probability of the principal's test is at most
\begin{align}
\label{EqnPosNullKnown}
\posNull \leq \nullApp \; \left \{ \frac{\altApp \altUtilGain -
  \utilLoss}{(\altApp - \nullApp) \utilLoss + \nullApp \altApp
  (\altUtilGain - \nullUtilGain)} \right \}.
\end{align}
\item \label[thm]{th:sharpness} For any $\threshold \in [0,1]$ such
  that $\bar \beta_0(\threshold) \in \big (0, \utilLoss/\nullUtilGain
  \big)$ and any pair $(\priorDist_0, \priorDist_1)$, there exists
  some $\priorNull \in [0,1]$ such that bound~\eqref{EqnPosNullKnown}
  is met with equality for the mixture prior~\eqref{EqnMixturePrior}.
\end{enumerate}
\end{theorem}
\noindent We prove this claim in~\Cref{AppProofThmBound}.  \\

\noindent {\bf{Remark:}} There is no loss of generality in requiring
that $\bar \beta_0(\threshold) \in \big(0, \; \utilLoss/\nullUtilGain
\big)$.  For a threshold such that $\bar \beta_0(\threshold) >
\utilLoss/\nullUtilGain$, even the ``worst agent'' with $\priorNull =
1$ would choose to \optin, leading to a Bayes FDR equal to one.

\paragraph{Implications for risk-neutral agents:} 
We can gain helpful intuition for the general
bound~\eqref{EqnPosNullKnown} by specializing it to a
simple-versus-simple test, a risk-neutral agent (with linear utility
function $\util(\wealth) = \wealth$) and constant reward $\constR$.
With these choices, we have $\nullApp = \beta_0(\threshold)$, $\altApp
= \beta_1(\threshold)$, $\UtilGain{j} = \constR$ for $j \in \{0,1 \}$,
and $\utilLoss = \cost$, and the bound~\eqref{EqnPosNullKnown} becomes
\begin{align*}
  \posNull & \leq \nullAppSimple \Bigg \{ \frac{\altAppSimple \constR - \cost}{
    (\altAppSimple - \nullAppSimple) \cost} \Bigg \} \; = \; \frac{\nullAppSimple
    \constR}{\cost} \Bigg \{ \frac{\altAppSimple \constR - \cost}{ (\altAppSimple
    - \nullAppSimple) \constR} \Bigg \}.
\end{align*}
Thus, we have recovered inequality~\eqref{EqnLinearBound}\Bone
in~\Cref{CorLinearBound} as a special case of the general
result~\eqref{EqnPosNullKnown}.

\paragraph{Role of assumptions:}  Returning
to the general setting, let us clarify how our assumptions are related
to the appearance of various terms in the
bound~\eqref{EqnPosNullKnown}.  The bound includes three
differences---all of which should be positive to yield a valid
upper bound on a probability.  First, from the non-trivial power
condition~\eqref{EqnNonTrivialCompPower}, the difference $\bar \beta_1(\threshold) -
\bar \beta_0(\threshold)$ is strictly positive; moreover, the
non-decreasing property of the utility and the assumption $\cost \geq
0$ imply that $\utilLoss \geq 0$. Second, it follows from the
stochastic monotonicity condition~\eqref{EqnStochMonotone} that
$\altUtilGain - \nullUtilGain \geq 0$.  Lastly, we need to verify that
\begin{align}
\label{EqnIdealOptIn}  
\bar \beta_1(\threshold) \altUtilGain - \utilLoss & \geq 0.
\end{align}
The validity of this inequality depends on the opt-in assumption
implicit in~\Cref{ThmBound}---namely, that there is some agent that
chose to \optin at threshold $\threshold$---along with the
utility-maximizing nature of agents.  These conditions ensure that the
``ideal agent''---meaning an agent with prior null probability
$\priorNull = 0$---would certainly \optin.  Combining these two
conditions allows us to establish inequality~\eqref{EqnIdealOptIn};
see the argument surrounding equation~\eqref{EqnTrickyAppendix}
in~\Cref{AppProofPriorBtwo} for the details.

\subsubsection{Conservative upper bounds}

If the principal lacks precise knowledge of the quantity $\nullApp$,
$\altApp$, $\nullUtilGain$, and $\altUtilGain$, she can still obtain a
conservative estimate of the guarantee~\eqref{EqnPosNullKnown} by
upper bounding these unknowns.  Recall that $\nullApp \leq \threshold$
by the super-uniformity condition~\eqref{EqnSuperUniform}; moreover, suppose
 that there is a known \emph{envelope function} $\threshold
\mapsto \kappa(\threshold)$ such that
\begin{subequations}
\begin{align}
\label{EqnKappaBound}
\sup_{\theta \in \Theta_1} \beta_\theta(\threshold) \leq
\kappa(\threshold) \qquad
\mbox{for each $\threshold \in [0, 1]$.}
\end{align}
For $j \in \{0,1 \}$, let us define
\begin{align}
\label{EqnRewardBounds}  
  \constR_j \defn \sup_{\theta \in \Theta_j} \Exs_\theta[\Reward]
  \quad \mbox{and} \quad \DelBar_j \defn \util(\Wealth_0 + \constR_j -
  \cost) - \util(\Wealth_0 - \cost),
\end{align}
\end{subequations}
where $\Exs_\theta$ denotes expectation under the distribution
$\Pdist_\theta$.  Finally, recall our previous definition
\mbox{$\utilLoss \defn \util(\Wealth_0) - \util(\Wealth_0 - \cost)$.}
\begin{corollary}
\label{CorPosNull}
For any decision threshold $\threshold$ such that $\nullApp <
\utilLoss / \nullUtilGain$, we have
\begin{align}
\label{EqnPosNullUnknown}
\posNull \leq \threshold \frac{\kappa(\tau) \DelBar_1 -
  \utilLoss}{(\kappa(\threshold) - \threshold) \utilLoss + \threshold
  \kappa(\tau) [\DelBar_1 - \DelBar_0] }.
\end{align}
\end{corollary}
\noindent See~\Cref{AppProofCorPosNull} for the proof of this claim.
Here we discuss its consequences for risk-neutral agents and
compare it to~\Cref{ThmBound}.

\paragraph{Implications for risk-neutral agents:}   Let us show
how~\Cref{CorPosNull}, when specialized to the linear utility function
$\util(\wealth) = \wealth$, constant reward $\constR$ and envelope
function $\kappa(\threshold) = 1$, recovers the bound~\Btwo from
equation~\eqref{EqnLinearBound} in~\Cref{CorLinearBound}.  With these
choices, we have $\DelBar_0 = \DelBar_1 = \constR$ and $\utilLoss =
\cost$.  Setting these values and $\kappa(\threshold) = 1$ into
inequality~\eqref{EqnPosNullUnknown}, we find that
\begin{align*}
\posNull & \leq \threshold \frac{\constR - \cost}{(1 - \threshold)
  \cost} \; = \; \threshold \frac{\constR}{\cost} \; \Biggr \{
\frac{\constR - \cost}{(1 - \threshold) \constR} \Biggr \},
\end{align*}
as claimed in inequality~\eqref{EqnLinearBound}\Btwo.

\paragraph{Comparing with\texorpdfstring{~\Cref{ThmBound}}{ Theorem 1}:}  It is worthwhile
comparing the upper bound~\eqref{EqnPosNullUnknown} with our earlier
upper bound~\eqref{EqnPosNullKnown} from~\Cref{ThmBound}.  For
discussion, let us refer to the latter as the \emph{known-$\beta$}
bound, and the former as the \emph{unknown-$\beta$} bound.  Suppose
that the rewards are deterministic, equal to separate values
$\constR_0$ and $\constR_1$ in the null and alternative cases,
respectively.  For such rewards, the two guarantees only differ in
that the potentially unknown values $(\beta_0(\threshold),
\beta_1(\threshold))$ are replaced by the corresponding upper bounds
$\big( \threshold, \kappa(\threshold) \big)$.  When the $p$-value is
actually uniform under the null, we have $\beta_0(\threshold) =
\threshold$, so nothing is lost in the upper bound.

The unknown-$\beta$ upper bound~\eqref{EqnPosNullUnknown} always holds
with $\kappa(\threshold) = 1$.  With this choice, the gap between the known and
unknown-$\beta$ cases should depend on the ``easiness'' of the testing
problem.  For instance, consider the case of a simple null $\Theta_0 =
\{0 \}$ versus the simple alternative $\Theta_1 = \{\theta_1\}$ for
Gaussian mean testing (see~\Cref{SecGaussMean} for the model set-up).
As $|\theta_1|$ increases, the power function
$\beta_{\theta_1}(\threshold)$ approaches $1$, so that the
unknown-$\beta$ bound should become a better approximation to the
known $\beta$-bound.  \Cref{FigRiskNeutral} provides an
illustration of this phenomenon in the special case of linear utility.

Otherwise, returning to general rewards, in order to compute the upper
bound~\eqref{EqnPosNullUnknown}, the principal requires uniform upper
bounds $\constR_0$ and $\constR_1$ of the reward means
$\Exs_\theta[\Reward]$ over the null and alternative spaces
(cf. equation~\eqref{EqnRewardBounds}). In the proof, these upper
bounds are combined with Jensen's equality, and the assumed concavity
and non-decreasing property of the utility function $\util$, to argue that
\begin{align}
  \label{EqnRiskPremium}
\Exs[\util(\Wealth_0 + R - \cost) \mid \theta \in \Theta_j] & \leq
\util(\Wealth_0 + \constR_j - \cost) \qquad \mbox{for $j \in \{0,
  1\}$.}
\end{align}
In the case of a simple-versus-simple hypothesis test---say with $\Theta_j =
\{\theta_j \}$ for $j \in \{0,1\}$---we have $\constR_j =
\Exs_{\theta_j}[\rfunc]$, so that the bounds~\eqref{EqnRiskPremium}
are sharp for a risk-neutral agent ($\util(\wealth) = \wealth)$.  For
risk-averse agents, the bounds~\eqref{EqnRiskPremium} become
progressively weaker as the degree of risk aversion increases.  This
gap can be understood via the lens of certainty equivalence (or risk
premia); for a risk-averse agent, the utility associated with
stochastic rewards is always less than the corresponding utility with
mean-matched deterministic rewards.


\subsubsection{A general form of prior elicitation}

The proof of~\Cref{th:bound_on_pos_null} involves an auxiliary result
that is of independent interest. In particular, a key analytical step
is to show that an agent's decision to \optin implies an upper bound
on the prior probability $\priorNull = \priorDist(\theta \in
\Theta_0)$ of the proposal being null. More precisely, we establish
the following fact:

\begin{proposition}[Prior elicitation]
  \label{PropPriorElicit}
Under the conditions of~\Cref{ThmBound}, fix some threshold
$\threshold$ such that \mbox{$\bar \beta_0(\threshold) \in (0, \;
  \utilLoss/\nullUtilGain)$.}  Then the prior null probability
$\priorNull$ of any agent who decides to \optin satisfies the upper
bounds
\begin{align}
\label{eq:bound_on_prior_null}
\priorNull & \; \stackrel{\Bone}{\leq} \; \frac{\altApp \altUtilGain -
  \utilLoss}{ \altApp \altUtilGain - \nullApp \nullUtilGain} \;
\stackrel{\Btwo}{\leq} \; \frac{\kappa(\threshold) \altUtilGain -
  \utilLoss}{ \kappa(\threshold) \altUtilGain - \threshold
  \nullUtilGain},
\end{align}
with the envelope function $\kappa(\threshold)$ defined in
equation~\eqref{EqnKappaBound}.
\end{proposition}
\noindent See~\Cref{AppProofPropPriorElicit} for the proof of this
claim. \\

\paragraph{Implications for risk-neutral agents:}
It is useful to study the implications of this claim for a
risk-neutral agent (with $\util(\wealth) = \wealth)$) and constant
reward $\constR$.  For the linear utility, we have $\UtilGain{0} =
\UtilGain{1} = \constR$ and $\utilLoss = \cost$.  For a
simple-versus-simple test, we have $\nullApp = \nullAppSimple$ and
$\altApp = \altAppSimple$. Substituting these choices into
equation~\eqref{eq:bound_on_prior_null}, we recover our previously
stated claim~\eqref{eq:linear_bound_on_prior_null} for risk-neutral
agents. \\

\Cref{PropPriorElicit} lies at the heart of the upper bounds on Bayes
FDR from~\Cref{ThmBound}.  In particular, we prove that for any fixed
threshold, the Bayes FDR is an increasing function of $\priorNull$, so
that any upper bound on $\priorNull$ induces an upper bound on the
Bayes FDR.  Thus, it is also intimately connected to the sharpness
guarantee stated in~\Cref{ThmBound}.  In particular, our proof shows
that the FDR bound in~\Cref{th:bound_on_pos_null} is met with equality
for any agent who chooses to \optin with a prior null probability $\priorNull$
that makes inequality~\eqref{eq:bound_on_prior_null}\Bone hold with
equality.  Any such agent is ``worst-case'' in the sense that its
prior null probability $\priorNull$ is as large as possible while
still being consistent with his decision to \optin at the threshold
$\threshold$.


\subsection{Staircase Sharpness with a Mixture of Agents}
\label{SecMain:staircase_sharpness}

In this section, we present a stronger sharpness result that is
naturally formulated in the setting of multiple agents.  Recall
that~\Cref{ThmBound}(b) provides a sharpness guarantee for the
single-agent case: there exists a prior null probability $\priorNull =
\priorDist(\theta \in \Theta_0)$ under which the upper
bound~\eqref{EqnPosNullKnown} on the Bayes FDR is exactly attained at
a single point.  The bound also applies to a mixture of different
agent types; \Cref{FigRiskNeutral} illustrates such a mixture with $K
= 2$ types of agent---say ``good'' versus ``bad''.  As shown in these
plots, there is some $\threshold$ at which only good agents choose to
\optin, and the upper bound matches the upper bound at this point
(consistent with~\Cref{ThmBound}(b)).  Moreover, the Bayes FDR is
actually very close to the bound at the threshold at which the bad
agents also choose to \optin. This phenomenon shown
in~\Cref{FigRiskNeutral} is, in fact, quite general: with a population
of agents, the exact Bayes FDR can be made \emph{arbitrarily close to
the known-$\beta$ bound at a countable number of points.} \\

To understand this fact, we begin by formalizing the meaning
of Bayes FDR for a mixture of agents.  Doing so requires two functions
that depend on a prior null probability $\priorNull$ and a threshold
$\threshold$.  In particular, the function
\begin{align*}
  \fdr(\priorNull; \threshold)&= \Prob(\theta \in \Theta_0 \mid
  \approves \mbox{ agents with } \priorNull \mbox{ under } \threshold)
\end{align*}
corresponds to the Bayes FDR induced by agents with prior null
$\priorNull$ who choose to \optin under threshold $\threshold$.  We
also define the indicator function
\begin{align*}
  \Ind(\priorNull ;\threshold) &= \begin{cases} 1 \qquad \mbox{if an
      agent with $\priorNull$ declares \optin under $\threshold$} \\ 0
    \qquad \mbox{otherwise}
    \end{cases},
\end{align*}
which tracks whether or not an agent with prior null $\priorNull$
chooses to \optin.

Now consider a mixture of $\numAgent$ types of agent, where each type
is characterized by a prior null probability $\priorNull^i \in [0,1]$
and a mixture proportion $\agentWeight{i} \in [0,1]$, normalized so
that $\sum_{i=1}^\numAgent \agentWeight{i} = 1$.  The exact Bayes FDR
under for this $\numAgent$-mixture, which we refer to as $\kfdr$, is
given by
\begin{subequations}
\begin{align}
  \label{EqnMixtureFDR}
\kfdr(\threshold) \defn \sum_{i=1}^\numAgent \frac{\agentWeight{i}
  \Ind(\priorNull^i;\threshold)}{\sum_{k=1}^K\agentWeight{k}
  \Ind(\priorNull^k;\threshold)} \fdr(\priorNull^i; \threshold),
\end{align}
assuming\footnote{If no agents choose to \optin, we set
$\kfdr(\threshold)$ to zero.} that at least one agent chooses to
\optin.  By definition, the quantity $\kfdr(\threshold)$ represents a
weighted average of the Bayes FDR for each agent type, where the
weights correspond to their proportions within the opt-in population.

We now show that it is possible to construct mixtures whose Bayes FDR
is $\epsilon$-close to the upper bound from~\Cref{ThmBound} at $K$
arbitrary points.  To focus on the essential ideas, we do so for
a simple-versus-simple hypothesis test, in which case the
known-$\beta$ bound~\eqref{EqnPosNullKnown} involves the function
\begin{align}
  \label{EqnBoundOne}
\Psi(\tau) \defn \nullAppSimple \; \left \{ \frac{\altAppSimple
  \altUtilGain - \utilLoss}{(\altAppSimple - \nullAppSimple) \utilLoss
  + \nullAppSimple \altAppSimple (\altUtilGain - \nullUtilGain)}
\right \}.
\end{align}
Furthermore, we assume that the $\numAgent$ types of agents share the
same initial wealth, trial cost, utility, and reward distributions,
differing only in their prior null probabilities $\priorNull^i$ and
mixture proportions $\agentWeight{i}$. Under these assumptions, the
same known-$\beta$ bound in~\eqref{EqnPosNullKnown} applies to each
agent type---that is, $\fdr(\priorNull^i;\threshold) \leq \Psi(\tau)$.
These inequalities imply that
\begin{align}
\kfdr(\threshold) & \leq \Psi(\threshold),
\end{align}
\end{subequations}
given the definition~\eqref{EqnMixtureFDR} of $\kfdr(\threshold)$ as a
convex combination.

In the following theorem, we establish that there are
$\numAgent$-mixtures of agents, as specified by some collection of
prior null probabilities $\{\priorNull^i\}_{i=1}^\numAgent$ and
weights $\{\agentWeight{i}\}_{i=1}^K$, for which the Bayes FDR
$\kfdr(\threshold)$ is arbitrarily close to $\Psi(\tau)$ at any
$\numAgent$ points.
\begin{theorem}[Staircase sharpness] 
\label{ThmStaircase}
Consider an increasing sequence of thresholds
$\{\threshold_i\}_{i=1}^\numAgent$ such that $\beta_0(\threshold_i)
\in \big (0, \utilLoss/\nullUtilGain \big)$ for all $i = 1, \ldots,
\numAgent$.  Then for any $\epsilon > 0$, there exists a
$\numAgent$-mixture of agents such that
\begin{align}
\label{EqnStaircase}
\Psi(\threshold_1) \; \stackrel{\Bone}{=} \; \kfdr(\threshold_1) \quad
\mbox{and} \quad \Psi(\threshold_i) \; \stackrel{\Btwo}{\leq} \;
\kfdr(\threshold_i) + \epsilon \qquad \mbox{for all $i \in
  \{2,\dots,\numAgent \}$}.
\end{align}
\end{theorem}
\noindent See~\Cref{AppProofThmStaircase} for the proof.\\

The sharpness guarantee~\eqref{EqnStaircase}---exact at a point and
$\epsilon$-close at $(K-1)$ points---is substantially stronger than
the result in~\Cref{th:sharpness}.  It shows that it is impossible to
improve~\Cref{ThmBound} in any substantive way without further
constraints on the agent mixtures. \\

To gain intuition for~\Cref{ThmStaircase}, it is helpful to
revisit the Gaussian mean testing example
from~\Cref{SecGaussMean}. Following the same setup, we consider
agents with a linear utility function, a cost $\cost = 10$, and a
constant reward $\constR = 25$.  For each choice of $K \in \{20, 40
\}$, we construct a $K$-mixture ensemble in which: (i) the agents have
prior null probabilities $\{\priorNull^i\}_{i=1}^K$ evenly spaced over
the interval $[0.02, 0.97]$; and (ii) the mixture proportions
$\{\agentWeight{i}\}_{i=1}^K$ are constructed to satisfy the condition
\begin{align*}
  \frac{\agentWeight{j}}{\sum_{i=1}^j \agentWeight{i}} = 0.99 \qquad
  \mbox{ for each $j \in \{2, \dots, K\}$.}
\end{align*}
This condition ensures that at each threshold at which a new type of
agents with a higher prior null probability \optin, they account for 99\% of
the opt-in population.

\begin{figure}[ht]
  \begin{center}
    \begin{tabular}{ccc}
      \widgraph{0.45\textwidth}{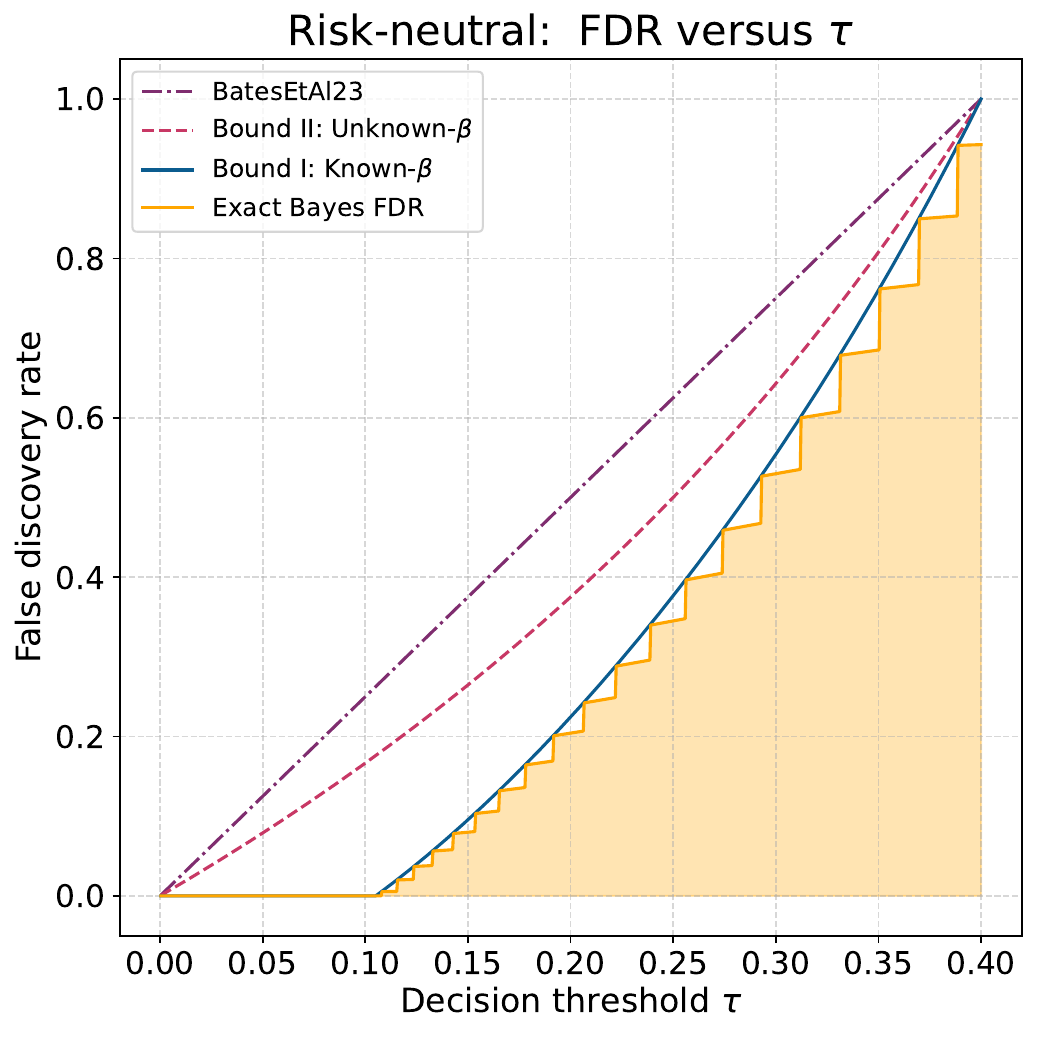}
      &&
      \widgraph{0.45\textwidth}{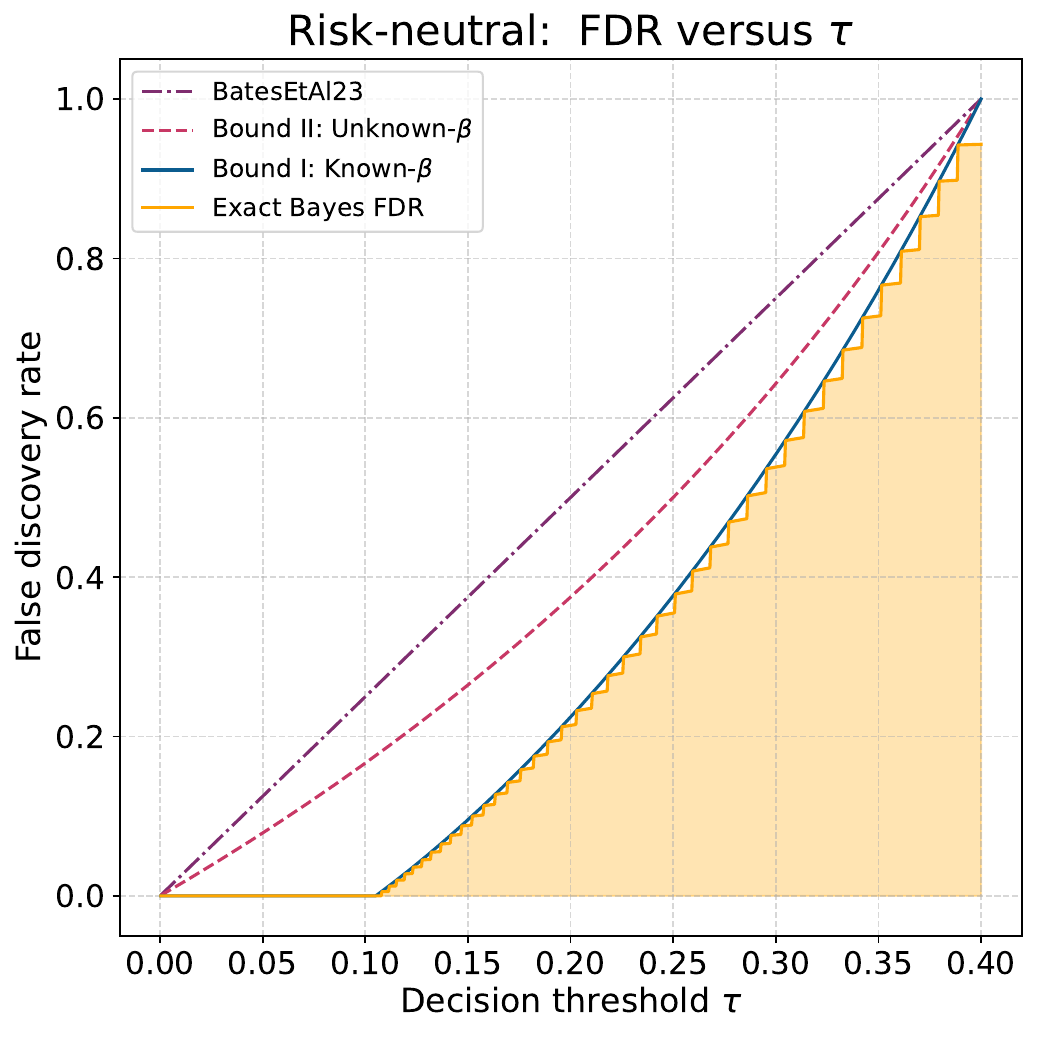}
      \\
      (a) Mixture of agents $K = 20$ && (b) Mixture of agents $K = 40$
    \end{tabular}
    \caption{Illustration of the staircase sharpness
      guarantee~\eqref{EqnStaircase} from~\Cref{ThmStaircase}.  Plots
      of the exact Bayes FDR; bounds (I) and (II) from
      equation~\eqref{EqnLinearBound}; and
      the~\citet{bates2023incentivetheoretic} bound as a function of
      the threshold choice $\threshold \in [0, \cost/\constR]$ for
      cost $\cost = 10$ and constant reward $\constR = 25$ with linear
      utility.  (a) A mixture of $K = 20$ types of agents.  (b) A
      mixture of $K = 40$ types of agents.}
    \label{FigStairCase}
  \end{center}
\end{figure}

In~\Cref{FigStairCase}, we plot the resulting
$\kfdr(\threshold)$ as the threshold $\threshold$ is varied, along
with bounds (I) and (II) from equation~\eqref{EqnLinearBound}; and
the~\citet{bates2023incentivetheoretic} bound.  Bound (I) is
equivalent to the bound defined by the function $\Psi$ from
equation~\eqref{EqnBoundOne}, and panels (a) and (b) correspond to $K
= 20$ and $K = 40$, respectively.
In panel (a), the exact Bayes FDR exhibits 20 step-like
transitions, each corresponding to the \optin of agents with
progressively higher prior null probabilities. At these transitions, the
exact Bayes FDR closely aligns with the known-$\beta$ bound, with the
maximum gap given by $\epsilon = 0.002$. In panel (b), we increase the
number of mixtures and observe a diminishing gap between the
known-$\beta$ bound and the exact Bayes FDR, with the maximum gap in
this case decreasing to $\epsilon = 0.0016$.

Our construction of the mixture ensemble in this example also offers
insight into~\Cref{ThmStaircase} and its connection to the sharpness
result in~\Cref{ThmBound}(b). Consider an increasing sequence of $K$
thresholds $\{\threshold_i\}_{i=1}^K$ and $K$ types of agents, where
agents of type $i$ have a prior null probability $\priorNull^i$ that
makes inequality~\eqref{eq:bound_on_prior_null}\Bone hold with
equality under $\threshold_i$. These agents represent the
``worst-case" agents that would \optin under $\threshold_i$. According
to~\Cref{ThmBound}(b), if only agents of type $i$ \optin under
$\threshold_i$, the known-$\beta$ bound is exact. With a mixture
ensemble, agents with prior null probabilities smaller than
$\priorNull^i$ will also \optin, thereby reducing the exact Bayes
FDR. \Cref{ThmStaircase} implies that if agents of type $i$ constitute
the vast majority of opt-in agents under $\threshold_i$, the known-$\beta$
bound remains nearly sharp, providing a highly accurate estimate of
the exact Bayes FDR. This highlights the ``worst-case" perspective
of~\Cref{ThmBound}: in the presence of information asymmetry, the
principal can use agents' incentives to guard against the worst-case
scenario. If these worst-case agents dominate the set of proposal
submissions, then the upper bound on the Bayes FDR established
in~\Cref{ThmBound}(a) is saturated.


\subsection{Principal's Utility with Maximin Guarantees}
\label{SecMain:maximin_optimality}

In our discussion thus far, we have focused purely on the agent's
utility. The principal might also wish to optimize some form of social
welfare, and in this section, we show how our theory leads to a
maximin guarantee for the principal's choice of threshold.

In particular, when dealing with an agent with a prior
$\priorDist$---call it a $\priorDist$-agent for short---the principal
might wish to maximize the societal benefits of approving effective
treatments ($\theta \in \Theta_1$) while minimizing the societal harms
of approving ineffective treatments ($\theta \in \Theta_0$). We view
the principal as taking an action $a \in \{\approves, \denies\}$, and
we now explicitly define the principal's utility as
\begin{align*}
\begin{cases}
    \omega_1 & \text{ if } \theta \in \Theta_1 \text{ and } a =
    \approves \\ -\omega_0 & \text{ if } \theta \in \Theta_0 \text{
      and } a = \approves \\ 0 & \text{ if } a = \denies
\end{cases},
\end{align*}
where $\omega_0$ and $\omega_1$ are positive weights.  Recall that the
principal's policy is to play action $\approves$ when $X \le \tau$.
The principal's average utility with threshold $\tau$ when confronted
with a $\priorDist$-agent is then given by
\begin{align*}
\Putil(\threshold; \priorDist) & = \begin{cases} \altweight
  \Prob(\theta \in \Theta_1 \mid \approves) - \nullweight \Prob(\theta
  \in \Theta_0 \mid \approves) & \mbox{when $\priorDist$-agent opts
    in} \\
    0 & \mbox{otherwise}
  \end{cases}.        
\end{align*}
  Now of course,
the principal does not know the prior distribution $\priorDist$;
consequently, it is natural to consider the worst-case
$\PutilBar(\threshold) \defn \inf_{\priorDist} \Putil(\threshold;
\priorDist)$.  It follows from the definition of $\Putil$ that for any
choice of threshold $\threshold$, we have $\PutilBar(\threshold) \leq
0$.  Accordingly, we say that a threshold choice $\threshold$ is
\emph{maximin optimal} if
\begin{align*}
\PutilBar(\threshold) = 0 \; = \; \max_{\tau' \in [0,1]}
\min_{\priorDist} \Putil(\tau', \priorDist).
\end{align*}
The following result characterizes the structure of maximin rules, in
terms of the upper bound~\eqref{EqnPosNullKnown}\Bone
from~\Cref{ThmBound}. 

To state the result, we need additional notation.  For any $\tau \in
(0,1)$, let $\widetilde \beta_0(\tau) = \sup_{\theta \in \Theta_0}
\Pdist_\theta(X \le \tau)$ and $\widetilde \beta_1(\tau) =
\sup_{\theta \in \Theta_1} \Pdist_\theta(X \le \tau)$. These are the
largest probabilities that the principal \approves when the threshold
is set to $\tau$ under the null and alternative, respectively. We
further assume that the distribution of $R$ depends only on whether or
not $\theta$ is in the set of nulls (i.e., on $\Ind(\theta \in
\Theta_0)$) so that $\Delta_0$ and $\Delta_1$ do not depend on
$\priorDist$.  Define the function
\begin{align*}
\widetilde \Psi(\threshold) \defn \widetilde \beta_0(\threshold) \;
\left \{ \frac{\widetilde \beta_1(\threshold) \altUtilGain -
  \utilLoss}{(\widetilde \beta_1(\threshold) - \widetilde
  \beta_0(\threshold)) \utilLoss + \widetilde \beta_1(\threshold)
  \widetilde \beta_0(\threshold) (\altUtilGain - \nullUtilGain)}
\right \},
\end{align*}
which~\Cref{ThmBound} guarantees to be a tight bound on the posterior
probability of null.  The following result states that it
characterizes the set of all maximin thresholds.
\begin{proposition}
\label{PropMaximin}
Under the above conditions and for given positive weights
$\nullweight$ and $\altweight$, a threshold $\threshold$ is maximin
optimal if and only if
\begin{align*}
\widetilde \Psi(\threshold) \leq \frac{\altweight}{\altweight + \nullweight}.
\end{align*}  
\end{proposition}
\noindent We provide the proof here, since it provides useful insight
into the connection with~\Cref{ThmBound}(b).
\begin{proof}
Note that $\sum_{j=0}^1 \Prob(\theta \in \Theta_j \mid \approves) =
1$, so for a $\priorDist$-agent who chooses to \optin, we have
\begin{align*}
  \Putil(\threshold, \priorDist) & = (\nullweight + \altweight) \;
  \Big \{ \frac{\altweight}{\nullweight + \altweight} - \Prob(\theta
  \in \Theta_0 \mid \approves) \Big \}.
\end{align*}
Since the utility is zero for $\priorDist$-agents who \optout,
we have
\begin{align*}
\PutilBar(\threshold) \defn \min \Big \{ 0, \inf_{\Qprob \in \mathcal Q^\tau} \Putil(\threshold,
\priorDist) \Big \} & = \min \Big \{ 0, (\nullweight + \altweight) \Big[
  \frac{\altweight}{\nullweight + \altweight} -
  \sup_{\priorDist \in \mathcal Q^\tau}
  \Prob(\theta \in \Theta_0 \mid \approves) \Big] \Big \},
\end{align*}
where $\mathcal Q^\tau$ is the set of all
distributions $\priorDist$ for which a $\priorDist$-agent chooses to \optin at
level $\threshold$.
But by the sharpness result
of~\Cref{ThmBound}(b), we have the equivalence
\mbox{$\sup_{\priorDist \in \mathcal Q^\tau} \Prob(\theta \in \Theta_0 \mid \approves) =
  \widetilde \Psi(\threshold)$,} so that
\begin{align*}
\PutilBar(\threshold) & = \min \Big \{ 0,
\; (\nullweight + \altweight) \Big[ \frac{\altweight}{\nullweight + \altweight} -
  \widetilde \Psi(\threshold) \Big] \Big \}.
\end{align*}
This representation shows that $\PutilBar(\threshold) = 0$ if and only
if $\widetilde \Psi(\threshold) \leq \frac{\altweight}{\nullweight +
  \altweight}$, as claimed.
\end{proof}

In short, the upper bound from~\Cref{ThmBound} implies that $\threshold$ such 
that $\widetilde \Psi(\tau) \le \altweight / (\nullweight +
  \altweight)$ is maximin optimal. In the other direction, 
  the sharpness result shows that for  
  any looser $\tau$, there exists some distribution with Bayes FDR larger
  than $\altweight / (\nullweight +
  \altweight)$, so such $\tau$ is not maximin optimal.
  
We introduce this maximin optimality result in part to facilitate
comparison with the findings of~\citet{tetenov2016economic}.  That
work shows that if agents have perfect information about $\theta$,
then taking $\tau$ as the agent cost divided by the agent reward is
maximin optimal. By contrast, our result shows that if agents have a
prior distribution rather than full information about $\theta$, a
stricter value of $\tau$ is typically required to achieve maximin
optimality.  In addition, that work also derives a maximin optimality
result for agents with prior distributions, but with a different
assumption about the principal's utility function. Roughly, under
certain assumptions about the underlying distributions and that the
principal's utility increases in effect size, the threshold of the agent
cost divided by reward is again shown to be maximin optimal.  Our result shows that with the utility function described above, maximin
optimality instead corresponds to controlling the Bayes FDR, and so
the sharp bound on the Bayes FDR yields an exact characterization of
maximin optimal rules.


\section{Numerical Experiments and Application to FDA Policy}
\label{SecApplication}

In previous sections, we present numerical results with two
objectives: comparing our upper bound on the Bayes FDR to prior work
and illustrating its sharpness.  In \Cref{SecApplication:normal}, we
shift focus to examine how the upper bound is influenced by the
agents' risk sensitivity and the stochasticity of the
reward. Revisiting the binary hypothesis test introduced
in~\Cref{SecGaussMean}, where the proposal quality $\theta$ represents
the mean of two normal distributions, we analyze the effects of
agents' risk preferences and reward variability. This analysis allows
us to explicitly demonstrate how agents' incentives can affect the
inference on the Bayes FDR under various type I error thresholds.

In \Cref{SecApplication:FDA}, we analyze the FDA's current clinical
trial approval policies within our principal-agent framework. Using
estimates of trial costs, profitability of drug development, and the
wealth and risk sensitivity of pharmaceutical companies, we assess the
false positive rate of the FDA's testing procedures.  Here, we show
how our framework leads to an improved understanding of how current
regulatory policies, which shape the financial incentives of
pharmaceutical companies, ultimately impact the quality of approved
drugs.

\subsection{Some qualitative phenomena}
\label{SecApplication:normal}

Recall that we consider the binary hypothesis test defined by the
parameter space $\Theta = \{0, \theta_1\}$ with the null set $\Theta_0
= \{0\}$ and the non-null set $\Theta_1 = \{\theta_1\}$ where
$\theta_1 > 0$, and the test statistic $Z \sim \mathcal{N}(\theta,
1)$. Throughout our experiments, we set $\theta_1 = 1$. Furthermore,
we consider a mixture ensemble of two types of agents: $10\%$ of the
agents are ``good'' with $\priorNull^g = 0.3$, and the remaining
$90\%$ are ``bad'' with $\priorNull^b = 0.8$. The cost of \optin is
$\cost = 10$, and both types of agents have an initial wealth of
$\Wealth_0 = 20$.

\subsubsection{Effects of risk aversion}

\begin{figure}[ht]
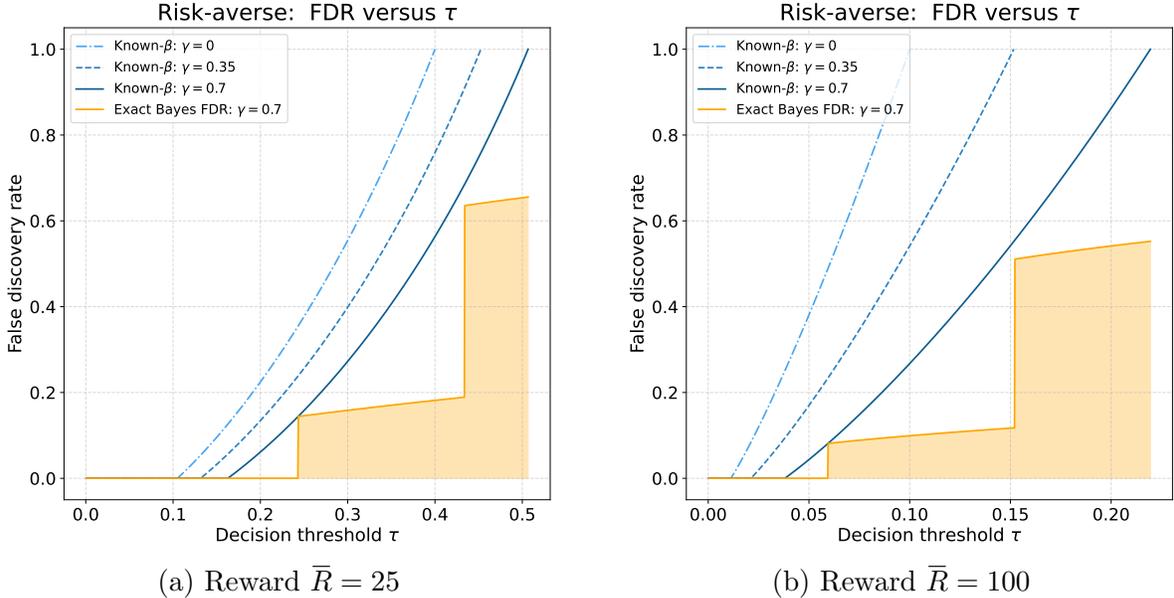

  \begin{center}
    \begin{tabular}{ccc}
      \widgraph{0.45\textwidth}{\figdir/fig_risk_averse_bound_r25} &&
      \widgraph{0.45\textwidth}{\figdir/fig_risk_averse_bound_r100}\\ (a)
      Reward $\constR = 25$ && (b) Reward $\constR = 100$
    \end{tabular}
    \caption{Upper bounds from equation~\eqref{EqnPosNullKnown}
      for agents with different degrees of risk aversion.}
    \label{FigRiskAverse}
  \end{center}
\end{figure}

In order to explore risk-sensitive behaviors, we perform experiments
using the class~\eqref{eq:risk_averse_utility} of utility functions with
constant relative risk
aversion~\citep{arrow1965riskaversion,PrattJohn1964}, indexed by some
parameter $\riskpar < 1$.  Recall that $\riskpar = 0$ corresponds to
risk neutrality, whereas $\riskpar \in (0, 1]$ yields a risk-averse
  loss.  Based on the analysis of~\citet{holt2002risk}, we pick
  $\riskpar = 0$ for \emph{risk-neutral} agents, $\riskpar = 0.35$ for
  \emph{slightly risk-averse} agents, and $\riskpar = 0.7$ for
  \emph{highly risk-averse} agents. For each choice of $\riskpar \in
  \{0, 0.35, 0.7\}$, we compute the known-$\beta$ bound as a function
  of the threshold $\threshold$. Assuming both types of agents are
  highly risk-averse, we also compute the exact Bayes FDR. These
  results are shown in~\Cref{FigRiskAverse}, where panel (a) uses a
  constant reward $\constR = 25$ and panel (b) uses a constant reward
  $\constR = 100$.

In both panels, we observe that the known-$\beta$ bounds for more
risk-averse utility functions reach a value of $1$ at significantly
larger threshold values, compared to the upper bound assuming
risk-neutral utility. As a direct consequence, the principal can
afford to set a looser $\threshold$ without incurring a high Bayes FDR
when the agents are risk-sensitive. However, if the principal
underestimates the agent's risk aversion---for instance, mistaking a
highly risk-averse agent for a risk-neutral one---then the principal
might set a threshold $\threshold$ that is too low for any agent to benefit from
opting in. Comparing panel (a) to panel (b), we observe that for a
fixed threshold $\threshold$, risk aversion introduces a greater
reduction in the known-$\beta$ bounds as the reward increases from
$25$ to $100$. This aligns with the concept of diminishing marginal
utility of wealth in economics. Since a more risk-averse agent has a
more concave utility function, their marginal utility decreases more
rapidly as wealth increases. As a result, a more risk-averse agent
derives a lower utility gain per unit of additional wealth, which
explains the widening gaps between the known-$\beta$ bounds. This
implies that for highly risk-averse agents, the Bayes FDR would not
increase significantly, even if the agent is offered a much larger
reward.

\subsubsection{Effects of reward stochasticity}

In this section, we investigate how random rewards influence the
known-$\beta$ bound in two key ways: (1) by increasing the gap between
the null and the non-null rewards, and (2) by introducing
stochasticity into both rewards.  Based on our previous discussion
(see~\eqref{EqnRiskPremium} and associated comments), stochastic
rewards exert a greater influence on agents with higher risk
aversion. Thus, we assume both types of agents are highly risk-averse
in the following simulation.

\begin{figure}[ht]
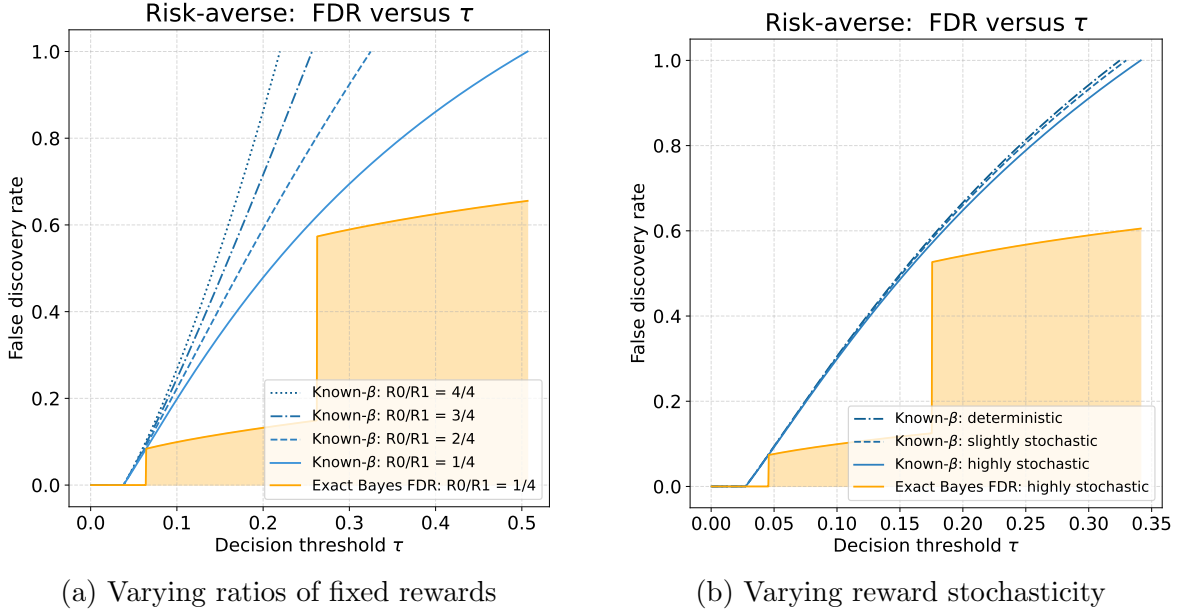

  \begin{center}
    \begin{tabular}{ccc}
      \widgraph{0.45\textwidth}{\figdir/fig_risk_averse_bound_rewardRatio} &&
      \widgraph{0.45\textwidth}{\figdir/fig_risk_averse_bound_randomReward} \\ (a)
      Varying ratios of fixed rewards && (b) Varying reward stochasticity
    \end{tabular}
    \caption{Upper bounds from equation~\eqref{EqnPosNullKnown} for
      slightly risk-averse agents with varying ratios and
      stochasticity of null and non-null rewards.}
    \label{FigRandomReward}
  \end{center}
\end{figure}

First, we examine the influence of the ratio between null and non-null
rewards. For illustrative purposes, we assume both rewards are
deterministic constants, with the non-null reward set to 100 and the
null reward selected from the set $\{100, 75, 50,
25\}$. In~\Cref{FigRandomReward}(a), we plot the known-$\beta$ bounds
for various reward ratios and the exact Bayes FDR assuming the null
reward is $25$. For a fixed threshold $\threshold$, the known-$\beta$
bound decreases as the null reward decreases. In other words, for a
fixed level of the Bayes FDR, the principal can set larger thresholds
as the null reward decreases. Notably, this change in threshold
$\threshold$ is not linear: the increase in $\threshold$ is much more
pronounced when the null reward drops from 50 to 25 than when it
decreases from 100 to 75. This suggests that even if the non-null
reward is 10 times the trial cost, agents who doubt the effectiveness
of their proposals will likely \optout, as their profits largely
depend on the null rewards. If the principal knows that the null
reward is substantially lower than the non-null reward, she can
leverage this information to achieve tighter control over the Bayes
FDR.

Next, we investigate the effects of reward stochasticity. We consider three levels of stochasticity: \emph{deterministic}, \emph{slightly stochastic}, and \emph{highly stochastic}. For the deterministic reward, we set the null reward to 50 and the non-null reward to 150. To model stochasticity, we use a truncated normal distribution. Let $\tnormal(\mu, \sigma, [a, b])$ denote a normal distribution with mean $\mu$ and variance $\sigma^2$ that lies within the interval $[a, b]$. For slightly stochastic rewards, we set:
\begin{align*}
    \reward_{slight} \mid \theta_0 \sim \tnormal(50, 25, [20, 80]) \quad \mbox{and} \quad \reward_{slight} \mid \theta_1 \sim \tnormal(150, 25, [120, 180]).
\end{align*}
For highly stochastic rewards, we set:
\begin{align*}
    \reward_{high} \mid \theta_0 \sim \tnormal(50, 35, [0, 100]) \quad \mbox{and} \quad \reward_{high} \mid \theta_1 \sim \tnormal(150, 35, [100, 200]).
\end{align*} 

In~\Cref{FigRandomReward}(b), we plot the known-$\beta$ bounds for
varying levels of reward stochasticity and the exact Bayes FDR
assuming highly stochastic rewards. As the reward becomes more
stochastic, the known-$\beta$ bound decreases, with the reduction
being more pronounced at larger thresholds.  Risk-averse agents prefer
certain outcomes over uncertain ones with the same expected value; see
equation~\eqref{EqnRiskPremium} and the associated discussion.
Consequently, if the principal knows that the agent is highly
risk-averse and the reward is highly volatile, the agent's decision to
\optin signals strong confidence in the proposal.


\subsection{Implications for FDA}
\label{SecApplication:FDA}

Next, we return to the motivating problem described in the
introduction---namely, the testing problem faced by the Food and Drug
Administration (FDA).  In particular, we study some implications of
our results using the range of costs, profits, and existing type I
error levels associated with clinical trials in the United States.

Let us begin with some relevant background.  The FDA mandates that
pharmaceutical companies conduct clinical trials to provide evidence
of the safety and efficacy of new drugs. The financial burden of
conducting these trials falls on the pharmaceutical companies. Based
on the trial results, the FDA then decides whether to grant
approval. If approved, the company can market the drug to the public
and generate profits.

The FDA retains some flexibility in the type and amount of evidence
that suffices for approval; see the
papers~\cite{bates2022principalagent,bates2023incentivetheoretic} for
a more detailed discussion of the FDA's approval guidelines. Following
\citet{bates2022principalagent}, we consider three simplified
statistical protocols that align with current FDA practice. We
evaluate (1) a {\it standard} protocol that requires a $p$-value below
0.05 in two independent trials; (2) a {\it modernized} protocol that
approves a drug if the $p$-value is less than 0.01 in a single trial;
(3) an {\it accelerated} protocol where two trials are conducted and
the drug is approved if either trial results in a $p$-value below
0.05. All $p$-values referenced above are two-sided.

The cost of conducting a Phase III clinical trial in the U.S. can vary significantly based on the therapeutic area, the number of patients, and the trial's complexity. A 2020 study by the Institute for Safe Medication Practices estimated the median cost of a pivotal clinical trial to be around \$48 million, with an interquartile range of \$20 million to \$102 million \citep{moore2020variation}. \citet{dimasi2016innovation} estimated an average out-of-pocket cost of \$255 million for a Phase III trial. Additionally, \citet{wouters2020estimated} provided a higher estimate of \$291 million. Note that these estimates pertain only to pivotal trials; the overall cost of drug development also includes expenses related to preclinical research, regulatory and legal costs, and investments in manufacturing and production. 
Given these estimates from the literature, we use an estimate of $\cost = \$200$ million as the cost of a trial in our
analysis.

Regarding the values of a company's reward, we note that the
profitability of a drug follows a long right-skewed distribution, with
substantial profits possible for the most commercially successful
drugs. For instance, Keytruda, a leading cancer immunotherapy,
generated annual sales of over \$25 billion in 2023. In a more typical
case, a successful new drug would generate annual sales of around
\$500 million to \$1 billion during its patent-protected
lifespan---see \citet{ledley2020profitability} for further analysis of
the profitability of major pharmaceutical companies.
Given substantial differences in drug profitability, we examine a
case where the average profit of a null drug is capped at \$800
million, {\it i.e.}, $R_0 = \$800$ million, and the average profit of
an effective drug $R_1$ ranges from \$1 billion to \$50 billion.

To model the risk aversion of pharmaceutical companies, we again
utilize the utility function specified in
equation~\eqref{eq:risk_averse_utility} with parameters $\riskpar \in
\{ 0, 0.35, 0.7 \}$ representing risk-neutral, slightly risk-averse,
and highly risk-averse companies, respectively. We assume all
companies have the same level of risk sensitivity. Turning to the
initial wealth, major pharmaceutical firms, such as Pfizer and Johnson
\& Johnson, typically generate revenues ranging from \$50 billion to
\$100 billion, while small to medium-sized companies have estimated
annual revenues between \$1 billion and \$10 billion. Given that large
companies often provide a broad range of medical services, the portion
of the budget allocated specifically for clinical trials is likely to
be significantly lower.  Accordingly, we adopt an estimate of \$5
billion for the initial wealth $\Wealth_0$.

\begin{table}[h!]
\centering
\begin{tabular}{|c|c|c|c|c|c|c|}
    \hline
    \multirow{2}{*}{\textbf{Protocol}} & \multirow{2}{*}{\textbf{Type I error $\boldsymbol{\threshold}$ (\%)}} & \multirow{2}{*}{\textbf{$\boldsymbol{R_1}$ (billion)}} & \multicolumn{4}{c|}{\textbf{Bounds on FDR (\%)}} \\
    \cline{4-7}
    &  &  & \textbf{$\boldsymbol{\threshold R_1 / \cost}$} & \textbf{Neutral} & \textbf{Slight} & \textbf{High} \\
    \hline
        \multirow{3}{*}{standard} & \multirow{3}{*}{0.0625} & \$1 &  0.3 &  0.25 & 0.24 & 0.23 \\
         &  & \$25 &  7.8 & 7.2 & 5 & 3.5\\
         &  & \$50 &  15.6 & 13.5  & 8.2 & 5.1 \\
    \hline
        \multirow{3}{*}{modernized} & \multirow{3}{*}{0.5} & \$1  & 2.5 & 2 & 1.93 & 1.87\\
         & & \$25  & 62.5 & 38.8 & 29.9 & 22.9 \\
         & & \$50  & n/a & 56 & 42.2 & 30.6 \\
    \hline
        \multirow{3}{*}{accelerated} & \multirow{3}{*}{5} & \$1  & 25 & 20 & 19.4 & 18.8 \\
        & & \$25  & n/a & 88.6 & 83.9 &  78.3 \\
        & & \$50  & n/a & 94 & 89.9 & 84.2 \\
    \hline
\end{tabular}
    \caption{Bounds on the fraction of false positives for companies with different levels of risk aversion, given three $p$-value thresholds $\threshold$, a cost of $\cost = \$200$ million, an initial wealth of $\Wealth_0 = \$5$ billion, a null reward of $R_0 = \$800$ million, and a non-null reward $R_1$ ranging from \$1 billion to \$50 billion.}
    \label{tab:FDA_example}
\end{table}

Based on these estimates, we report the FDR bounds derived in
\citet{bates2023incentivetheoretic} and implied by our
\Cref{CorPosNull} in \Cref{tab:FDA_example}. To apply
\Cref{CorPosNull}, we suppose $\kappa = 1$. We also assume that the
drug will receive a reward of $R_1$ even if it might be ineffective
when applying the result in \citereset
\citet{bates2023incentivetheoretic}. A value of ``n/a" indicates that
the FDR bound is larger than 1, in which case the theory does not
indicate reasonable control of FDR. Compared to
\citereset\citet{bates2023incentivetheoretic}, our result suggests a
much lower fraction of false positives for risk-neutral agents across
the three protocols. In regimes where \citereset
\citet{bates2023incentivetheoretic} suggests a false discovery rate
above 1, our bound still provides a nontrivial bound on the FDR. When
the companies are risk-sensitive, we see a significant reduction in
FDR especially for highly profitable drugs that can earn \$25 billion
or more. This analysis suggests that the FDA might consider loosening
the significance level for less lucrative drugs.

It is crucial to note that \Cref{tab:FDA_example} is based on our
simplified theoretical model of how pharmaceutical companies interact
with the FDA. In reality, the utility functions of these companies
could be more complex and not fully captured by
\cref{eq:risk_averse_utility}. Our analysis seeks to clarify how false
discovery rates in clinical trials might be affected by the incentives
of various stakeholders. It provides guidance on how factors such as
risk sensitivity and variations in rewards together with the nominal
type I error rate impact the average quality of the approved drugs.


\section{Discussion}
\label{SecDiscussion}

In this paper, we have studied the problem of principal-agent
hypothesis testing, a specific instance of statistical
decision-making involving strategic agents.  This problem---and others of
its type---establishes connections between statistical inference and
incentive alignment in economic theory.  A fundamental challenge in
principal-agent testing lies in the fact that the quality of the data seen by the
principal---and hence the false discovery rate (FDR) associated with
her ultimate decision-making---is impacted by the agents' incentives. These incentives
are determined by the interplay of their utility functions, private
prior information, and the probabilistic structure of the experiment and rewards.

Among the main results of this paper are a general upper bound on the
Bayes FDR of the principal's test (\Cref{ThmBound}) and the ``staircase
sharpness'' property it satisfies (\Cref{ThmStaircase}).  Both results hold for general utility functions and stochastic
rewards, making them relevant to a broad class of decision models and
environments. An important implication of these two findings is that,
without making further assumptions on agent behavior and priors, it is
impossible to guarantee any tighter control on the Bayes FDR.  Given
this insight, one promising future direction is to explore different
forms of information asymmetry, and/or mechanisms by which the principal might enhance test performance by leveraging additional
knowledge about the distribution of agents.

Finally, aside from the specific technical contributions of this
paper, at a higher level, one interesting takeaway message is the connection between information asymmetry in statistical inference with
strategic agents and mechanisms for prior elicitation.  In the
current work, since the principal is unaware of the agents' priors, inferential conclusions about the FDR can only be drawn if the
agents' actions are information-revealing.  Thus, as formalized in our
analysis, the performance of statistical protocols hinges on their
ability to extract this information and on how this
information influences statistical inference.  In future work,
it would be interesting to broaden the range of possible mechanisms
for information elicitation---for instance through the use of multiple
contracts (as opposed to the single-contract model studied here) or
via sequential forms of interaction, in which the principal and agent
engage in multiple rounds of communication.

\subsection*{Acknowledgements}
The authors thank Alberto Abadie, Isaiah Andrews, Lihua Lei, Whitney
Newey, Ashesh Rambachan, and Davide Viviano for their comments on this
work.  In addition, this work was partially supported by
NSF-DMS-2413875 to SB and MJW, and the Cecil H. Green Chair to MJW.

\AtNextBibliography{\small} \printbibliography


\appendix

\section{Proofs}
\label{secProofs}

In this appendix, we collect together the proofs of our results.

\subsection{Proof of\texorpdfstring{~\Cref{PropPriorElicit}}{ Proposition 1}}
\label{AppProofPropPriorElicit}

We begin by proving the upper bounds on the prior null probability
$\priorNull = \priorDist(\theta \in \Theta_0)$
from~\Cref{PropPriorElicit}.  We split our proof into two parts, one
for each inequality in equation~\eqref{eq:bound_on_prior_null}.

\subsubsection{Proof of inequality\texorpdfstring{~\eqref{eq:bound_on_prior_null}\Bone}{ (20) (I)}}
\label{AppProofPriorBone}

By assumption, the agent with prior $\priorDist$, initial wealth
$\Wealth_0$ and utility function $\util$ has chosen to \optin with cost $\cost$. Recall that the utility-maximizing
nature of the agent ensures that we must have
\begin{align}
\label{EqnParticipate}      
  \E \Big[\util(\WealthAfter) \Big] & \geq \util(\Wealth_0),
\end{align}
which we refer to as the participation condition.  To be clear, the
expectation on the left-hand side is taken over $\theta \sim
\priorDist$, as well as the $p$-value $(X \mid \theta) \sim
\Pdist_\theta$ and stochastic reward $\rfunc$, where the pair $(X,
\rfunc)$ are conditionally independent given $\theta$.

Our next step is to derive an explicit expression for this expectation
that exposes its dependence on the prior probability $\priorNull$; in
particular, we show that it can be written as a linear function
of $\priorNull$ with negative slope. To ease notation, we adopt
the shorthand $\bar \beta_j = \bar \beta_j(\threshold)$ for $j \in \{0,1 \}$,
and recall the definitions
\begin{subequations}
  \label{EqnRedefine}
  \begin{align}
\label{EqnRedefineLoss}    
  \utilLoss & \defn \util(\Wealth_0) - \util(\Wealth_0 - \cost), \quad
  \mbox{and} \\
\label{EqnRedefineGain}  
\UtilGain{j} & \defn \E[\util(\WealthGain) \mid \theta \in \Theta_j] -
\util(\Wealth_0 - \cost) \quad \mbox{for $j \in \{0,1 \}$.}
\end{align}
\end{subequations}
In terms of this shorthand, we claim the equivalence
\begin{subequations}
  \begin{align}
\label{EqnLinEquivalence}    
\E[\util(\WealthAfter)] = \LinFun(\priorNull) \defn \priorNull \big[
  \nullAppAbbrev \nullUtilGain - \altAppAbbrev \altUtilGain \big] +
\big[ \util(\WealthLose) + \altAppAbbrev \altUtilGain \big],
  \end{align}
where the linear function $\priorNull \mapsto \LinFun(\priorNull)$ has
a negative slope---that is
\begin{align}
\label{EqnNegativeSlope}  
\nullAppAbbrev \nullUtilGain - \altAppAbbrev \altUtilGain < 0.
\end{align}
\end{subequations}

Taking~\eqref{EqnLinEquivalence} and~\eqref{EqnNegativeSlope} as given for the moment, let us
establish the claimed upper bound on $\priorNull$.  Combining the
participation condition~\eqref{EqnParticipate} with the
equivalence~\eqref{EqnLinEquivalence} yields the inequality
\begin{align*}
  \LinFun(\priorNull) \defn \priorNull \big[ \nullAppAbbrev
    \nullUtilGain - \altAppAbbrev \altUtilGain \big] + \big[
    \util(\WealthLose) + \altAppAbbrev \altUtilGain \big] & \geq
  \util(\Wealth_0).
\end{align*}
Given the negative slope~\eqref{EqnNegativeSlope}, we can divide both
sides by the slope and reverse the inequality sign.  Doing so and
re-arranging while making use of the definitions~\eqref{EqnRedefine}
yields the upper bound
\begin{align*}
\priorNull & \leq \frac{ \util(\WealthLose) + \altAppAbbrev
  \altUtilGain - \util(\Wealth_0)}{\altAppAbbrev \altUtilGain -
  \nullAppAbbrev \nullUtilGain} \; = \; \frac{ \altAppAbbrev
  \altUtilGain - \utilLoss }{\altAppAbbrev \altUtilGain -
  \nullAppAbbrev \nullUtilGain},
\end{align*}
which establishes inequality~\Bone from
equation~\eqref{eq:bound_on_prior_null}. \\

\noindent It remains to prove our two auxiliary
claims~\eqref{EqnLinEquivalence} and~\eqref{EqnNegativeSlope}.

\paragraph{Proof of claim\texorpdfstring{~\eqref{EqnLinEquivalence}}{ (34a)}:}

Introducing the shorthand $Y \defn \util(\WealthAfter)$, we begin by
writing
\begin{align}
\Exs[Y] & = \priorNull \Exs[Y \mid \theta \in \Theta_0] + (1 -
\priorNull) \Exs[Y \mid \theta \in \Theta_1] \notag \\
\label{EqnInitialDecomp}  
& = \priorNull \Big \{ \Exs[Y \mid \theta \in \Theta_0] - \Exs[Y \mid
  \theta \in \Theta_1] \Big \} + \Exs[Y \mid \theta \in \Theta_1].
\end{align}
Next, we compute the two conditional expectations appearing in
equation~\eqref{EqnInitialDecomp}. Define the binary random variable
$Z \defn \Ind(X \leq \threshold) \in \{0,1\}$, and observe that
$\beta_j(\threshold) = \Exs[Z \mid \theta \in \Theta_j]$ for $j \in
\{0,1 \}$.  Thus, we can write
\begin{subequations}
\begin{align}
\Exs[Y \mid \theta \in \Theta_0] & = \nullAppAbbrev \Exs[Y \mid
  \theta \in \Theta_0, Z = 1] + \big(1 - \nullAppAbbrev \big) \;
\Exs[Y \mid \theta \in \Theta_0, Z = 0] \nonumber \\
& \stackrel{(i)}{=} \nullAppAbbrev \Exs_R[\util(\Wealth_0 +
  \rfunc - \cost) \mid \theta \in \Theta_0] + \big(1 -
\nullAppAbbrev \big) \Exs_R[\util(\Wealth_0 - \cost) \mid \theta
  \in \Theta_0] \notag \\
\label{EqnCondOne}
& \stackrel{(ii)}{=} \nullAppAbbrev \underbrace{ \big \{
  \Exs_R[\util(\Wealth_0 + \rfunc - \cost) \mid \theta \in \Theta_0] -
  \util(\Wealth_0 - \cost) \big \}}_{\equiv \UtilGain{0}} +
\util(\Wealth_0 - \cost)
\end{align}
where step (i) uses the fact that the reward $\rfunc$ and decision $Z$
are conditionally independent given $\theta$; and step (ii) uses the
definition~\eqref{EqnRedefineGain} of $\UtilGain{0}$.  A similar
argument yields
\begin{align}
  \Exs[Y \mid \theta \in \Theta_1] & = \altAppAbbrev
  \Exs_R[\util(\Wealth_0 + \rfunc - \cost) \mid \theta \in \Theta_1] +
  \big(1 - \altAppAbbrev \big) \; \util(\Wealth_0 - \cost) \notag \\
\label{EqnCondTwo}  
  & = \altAppAbbrev \underbrace{\big \{ \Exs_R[\util(\Wealth_0 +
    \rfunc - \cost) \mid \theta \in \Theta_1] - \util(\Wealth_0 -
  \cost) \big \}}_{\equiv \UtilGain{1}} + \util(\Wealth_0 - \cost)
\end{align}
\end{subequations}
Finally, combining equations~\eqref{EqnCondOne} and~\eqref{EqnCondTwo}
with the initial decomposition~\eqref{EqnInitialDecomp} yields
\begin{align*}
\Exs[Y] & = \priorNull \big [ \nullAppAbbrev \UtilGain{0} -
\altAppAbbrev \UtilGain{1} \big ] + \altAppAbbrev
\UtilGain{1} + \util(\Wealth_0 - \cost),
\end{align*}
which completes the proof of the claim~\eqref{EqnLinEquivalence}.


\paragraph{Proof of claim\texorpdfstring{~\eqref{EqnNegativeSlope}}{ (34b)}:}

Recall that $\nullAppAbbrev < \altAppAbbrev$ by assumption.  As a consequence, in
order to show that $\nullAppAbbrev \nullUtilGain < \altAppAbbrev \altUtilGain$, it
suffices to prove that $\nullUtilGain \leq \altUtilGain$.  From the
definition~\eqref{EqnRedefineGain} of $\nullUtilGain$ and
$\altUtilGain$, it is equivalent to show
\begin{align*}
  \E[\util(\WealthGain) \mid \theta \in \Theta_0] & \leq
  \E[\util(\WealthGain) \mid \theta \in \Theta_1].
\end{align*}
This inequality is a consequence of the combination of the stochastic
monotonicity condition~\eqref{EqnStochMonotone} and the assumption
that the utility function $\util$ is non-decreasing.


\newcommand{\crefsection}[2]{%
  \section{\texorpdfstring{\Cref{#1}}{#2}}%
}
\newcommand{\eqrefsection}[2]{%
  \section{\texorpdfstring{\eqref{#1}}{#2}}%
}

\subsubsection{Proof of inequality\texorpdfstring{~\eqref{eq:bound_on_prior_null}~\Btwo}{ (20) (II)}}
\label{AppProofPriorBtwo}

We have proved inequality~\eqref{eq:bound_on_prior_null}\Bone.  In a
compact form, it can be written as $\priorNull \leq
\Psi(\nullAppAbbrev, \altAppAbbrev)$, where the bivariate function
$\Psi$ is given by
\begin{align*}
\Psi(\nullAppAbbrev, \altAppAbbrev) & \defn \frac{ \altAppAbbrev
  \altUtilGain - \utilLoss }{\altAppAbbrev \altUtilGain -
  \nullAppAbbrev \nullUtilGain}.
\end{align*}
In terms of the function $\Psi$, inequality~\Btwo is equivalent to
$\Psi(\nullAppAbbrev, \altAppAbbrev) \leq \Psi(\threshold, \kappa)$
for any $\kappa \in [\altAppAbbrev, 1]$.  Thus, it suffices to show
that the function $\Psi$ is increasing in each of its arguments.

By inspection, for each fixed $\altAppAbbrev \in [\threshold,1]$, the
function $\nullAppAbbrev \mapsto \Psi(\nullAppAbbrev, \altAppAbbrev)$
is increasing.  Turning to its second argument, it suffices to show
that for each fixed $\nullAppAbbrev \in [0, \threshold]$, the function
\begin{align*}
  f(\altAppAbbrev) \defn \log \Psi(\nullAppAbbrev, \altAppAbbrev) \; =
  \; \log(\altAppAbbrev \altUtilGain - \utilLoss) - \log(\altAppAbbrev
  \altUtilGain - \nullAppAbbrev \nullUtilGain)
\end{align*}
has a non-negative derivative for $\altAppAbbrev \in (\threshold, 1]$.
  Computing the derivative yields
\begin{align*}
f'(\altAppAbbrev) & = \frac{\altUtilGain}{\altAppAbbrev \altUtilGain -
  \utilLoss} - \frac{\altUtilGain}{\altAppAbbrev \altUtilGain -
  \nullAppAbbrev \nullUtilGain} \; = \; \altUtilGain \Big \{
\frac{\utilLoss - \nullAppAbbrev \nullUtilGain}{ (\altAppAbbrev
  \altUtilGain - \utilLoss) \; (\altAppAbbrev \altUtilGain -
  \nullAppAbbrev \nullUtilGain)} \Big \}.
\end{align*}
It suffices to show each of the four terms appearing on the right-hand
side are non-negative.  It is immediate that $\altUtilGain \geq 0$,
and we have $\utilLoss - \nullAppAbbrev \nullUtilGain \geq 0$ by
assumption.  The bound~\eqref{EqnNegativeSlope} from the proof of
inequality \Bone is equivalently stated as $\altAppAbbrev \altUtilGain
- \nullAppAbbrev \nullUtilGain > 0$.

The only remaining step is to show that
\begin{align}
\label{EqnTrickyAppendix}
\altAppAbbrev \altUtilGain - \utilLoss & \geq 0.
\end{align}
Note that the statement of~\Cref{PropPriorElicit} is predicated upon
the principal who declares the threshold $\threshold$ observes an agent that chooses to \optin.  From the participation
condition~\eqref{EqnParticipate}, for an agent with prior $\priorNull$
who decides to \optin, we must have $\LinFun(\priorNull) \geq \util(\Wealth_0)$,
where the linear function $\LinFun$ was defined in
equation~\eqref{EqnLinEquivalence}.

If any agent decides to \optin, then certainly the ``ideal'' agent with prior
null $\priorNull = 0$ would also \optin.   Consequently, we must
have
\begin{align*}
\LinFun(0) = \util(\WealthLose) + \altAppAbbrev \altUtilGain & \geq
\util(\Wealth_0)
\end{align*}
Since $\utilLoss = \util(\Wealth_0) - \util(\WealthLose)$ by
definition, we have established the claim~\eqref{EqnTrickyAppendix}.


\subsection{Proof of\texorpdfstring{~\Cref{ThmBound}}{ Theorem 1}}
\label{AppProofThmBound}

We now turn to the proof of~\Cref{ThmBound}, splitting our argument
into the two parts given in the statement itself.

\subsubsection{Proof of part (a):  Upper bound on Bayes FDR}
\label{AppProofThmA}

We prove the claimed upper bound by expressing the Bayes FDR as a
function of the prior null probability $\priorNull = \priorDist(\theta
\in \Theta_0)$, and then showing that the Bayes FDR is increasing
as a function of $\priorNull$.  These facts allow us to translate the
upper bound on $\priorNull$ from~\Cref{PropPriorElicit} into an upper
bound on the Bayes FDR.

More precisely, our proof hinges on the following two auxiliary claims:
the Bayes FDR can be written as $\posNull = \varphi(\priorNull)$,
where
\begin{subequations}
  \begin{align}
\label{EqnFDRPhi}    
\varphi(\priorNull) & \defn \frac{\priorNull
  \nullApp}{\priorNull(\nullApp - \altApp) + \altApp},
\end{align}
and moreover, we have
\begin{align}
\label{EqnMonotonePhi}  
\varphi'(\priorNull) \geq 0 \qquad \mbox{for all $\priorNull \in [0,1]$, so
  that $\varphi$ is an increasing function.}
\end{align}
\end{subequations}

We return to prove these two claims momentarily; taking them as given
for the moment, let us complete the proof of the claimed upper
bound~\eqref{EqnPosNullKnown} from the theorem
statement. Introduce the simpler notation $\bar \beta_j \equiv
\bar \beta_j(\threshold)$ for $j = 0, 1$, along with with the shorthand
\mbox{$\pUpper \defn \frac{\nullAppAbbrev \altUtilGain - \utilLoss}{ \altAppAbbrev
    \altUtilGain - \nullAppAbbrev
    \nullUtilGain}$} for the upper bound
on $\priorNull$ from~\Cref{PropPriorElicit}.  Now we can argue that
\begin{align*}
  \posNull \; \stackrel{(i)}{=} \; \varphi(\priorNull) \;
  \stackrel{(ii)}{\leq} \; \varphi(\pUpper) & = \frac{\pUpper
    \nullAppAbbrev}{\pUpper (\nullAppAbbrev - \altAppAbbrev) + \altAppAbbrev} \\
  &\stackrel{(iii)}{=} \frac{(\altAppAbbrev \altUtilGain - \utilLoss) \nullAppAbbrev}{(\altAppAbbrev \altUtilGain - \utilLoss)(\nullAppAbbrev - \altAppAbbrev) + (\altAppAbbrev
    \altUtilGain - \nullAppAbbrev
    \nullUtilGain)\altAppAbbrev}\\
  & = \nullAppAbbrev \left\{\frac{ 
 \altAppAbbrev
    \altUtilGain - \utilLoss}{( \altAppAbbrev - \nullAppAbbrev) \utilLoss + \nullAppAbbrev
    \altAppAbbrev (\altUtilGain - \nullUtilGain)}\right\},
\end{align*}
where step (i) follows from the equivalence~\eqref{EqnFDRPhi}, whereas
step (ii) follows from the increasing property~\eqref{EqnMonotonePhi},
and the upper bound $\priorNull \leq \pUpper$
from~\Cref{PropPriorElicit}. In step (iii), we plug in $\pUpper$ and multiply both the numerator and denominator by $\beta_1
    \altUtilGain - \nullAppAbbrev
    \nullUtilGain$. This completes the proof of the
bound~\eqref{EqnPosNullKnown}. \\

\noindent It remains to establish our two auxiliary claims.

\paragraph{Proof of claim\texorpdfstring{~\eqref{EqnFDRPhi}}{ (38a)}:}
Using the definition of the Bayes FDR, we have
\begin{align*}
\posNull & = \frac{\Prob(\approves \mid \theta \in
  \Theta_0)\Prob(\theta \in \Theta_0)}{\Prob(\approves)} \\
& = \frac{\Prob(\approves \mid \theta \in \Theta_0) \Prob(\theta \in
  \Theta_0)}{\Prob(\approves \mid \theta \in \Theta_0) \Prob(\theta
  \in \Theta_0) + \Prob(\approves \mid \theta \in \Theta_1)
  \Prob(\theta \in \Theta_1)} \\
& = \frac{\priorNull \nullAppAbbrev}{\priorNull \nullAppAbbrev + (1-\priorNull)
  \altAppAbbrev} \; \equiv \; \varphi(\priorNull),
\end{align*}
as claimed.

\paragraph{Proof of the claim\texorpdfstring{~\eqref{EqnMonotonePhi}}{ (38b)}:}
If $\nullAppAbbrev = 0$, then we have $\varphi(\priorNull) = 0$ for all
$\priorNull$, so that the claim follows immediately.  Otherwise, we
may take both $\nullAppAbbrev$ and $\altAppAbbrev$ to be positive, and taking
derivatives of $\varphi$ via the chain rule yields
\begin{align*}
\varphi'(\priorNull) & = \frac{[\priorNull(\nullAppAbbrev - \altAppAbbrev) +
    \altAppAbbrev] \nullAppAbbrev - (\nullAppAbbrev - \altAppAbbrev)\priorNull
  \nullAppAbbrev}{[\priorNull(\nullAppAbbrev - \altAppAbbrev) + \altAppAbbrev]^2} \; = \;
\frac{\nullAppAbbrev \altAppAbbrev}{[\priorNull(\nullAppAbbrev - \altAppAbbrev) + \altAppAbbrev]^2},
\end{align*}
which is strictly positive by inspection.  Thus, we have established
the claim~\eqref{EqnMonotonePhi}.


\subsubsection{Proof of part (b): Sharpness of the bound} 
\label{AppProofThmSharp}

Fix a pair of distributions $\{\priorDist_j\}_{j=0}^1$ and consider
the mixture prior $\priorDist = \priorNull \priorDist_0 +
(1-\priorNull)\priorDist_1$.  With the distributions $\priorDist_0$
and $\priorDist_1$ fixed, all of the functions $\nullApp$, $\altApp$,
$\nullUtilGain$, $\altUtilGain$ are also fixed and do not depend on
$\priorNull$. Fix some threshold $\threshold \geq 0$ such that
$\nullApp < \utilLoss/\nullUtilGain$, and introduce the shorthand
notation $\bar \beta_j = \bar \beta_j(\threshold)$ for $j = 0, 1$.
Consider an agent that behaves rationally with respect to some utility
function $\util$.  From the proof of~\Cref{PropPriorElicit}, their
decision boundary for an \optin decision is determined by the
inequality
\begin{align*}
  \priorNull \big[ \nullAppAbbrev \nullUtilGain - \altAppAbbrev
    \altUtilGain \big] + \big[ \util(\WealthLose) + \altAppAbbrev
    \altUtilGain \big] & \geq \util(\Wealth_0),
\end{align*}
and moreover, we previously established that $\nullAppAbbrev
\nullUtilGain - \altAppAbbrev \altUtilGain < 0$.

We claim that by an appropriate choice of their prior $\priorDist$,
and hence their prior null probability $\priorNull = \priorDist[\theta
  \in \Theta_0]$, we can ensure that their initial utility
$\util(\Wealth_0)$ is \emph{equal} to their expected utility for an
\optin decision---that is $\util(\Wealth_0) =
\E[\util(\WealthAfter)]$.

From the proof of~\Cref{PropPriorElicit}, it follows that the prior
null probability $\priorNull$ of the agent satisfies
\begin{align*}
\priorNull \big[ \nullAppAbbrev \nullUtilGain - \altAppAbbrev
  \altUtilGain \big] &= \utilLoss - \altAppAbbrev \altUtilGain,
\end{align*}
whence \mbox{$\priorNull = \frac{\altAppAbbrev \altUtilGain -
    \utilLoss}{\altAppAbbrev \altUtilGain - \nullAppAbbrev
    \nullUtilGain}$.}  Since the bound on $\priorNull$ is achieved
with equality, substituting this value into the function $\varphi$
from the proof of~\Cref{th:bound_on_pos_null} yields the desired
result. (The assumption $\nullApp < \utilLoss/\nullUtilGain$ ensures
that $\priorNull \in [0,1]$.)


\subsection{Proof of\texorpdfstring{~\Cref{CorPosNull}}{ Corollary 2}}
\label{AppProofCorPosNull} 

For a fixed threshold $\threshold$, we use the shorthand notation
$\bar \beta_j = \bar \beta_j(\threshold)$ for $j \in \{0,1 \}$, and $\kappa
\defn \kappa(\threshold)$.  Define the function
\begin{align*}
  \Phi(\nullAppAbbrev, \altAppAbbrev, \UtilGain{0}, \UtilGain{1}) & \defn
  \log(\nullAppAbbrev) + \log \big(\altAppAbbrev \UtilGain{1} - \utilLoss \big) -
  \log \big[ (\altAppAbbrev - \nullAppAbbrev) \utilLoss + \nullAppAbbrev \altAppAbbrev
    (\UtilGain{1} - \UtilGain{0}) \big].
\end{align*}
The statement of~\Cref{th:bound_on_pos_null} is equivalent to $\log \posNull
\leq \Phi(\nullAppAbbrev, \altAppAbbrev, \UtilGain{0}, \UtilGain{1})$, and the
statement of~\Cref{CorPosNull} is equivalent to
\begin{subequations}
\begin{align}
\label{EqnCleanRestate}  
\Phi(\nullAppAbbrev, \altAppAbbrev, \UtilGain{0}, \UtilGain{1}) \leq
\Phi(\threshold, \kappa, \DelBar_0, \DelBar_1).
\end{align}
By assumption, we have $\nullAppAbbrev \leq \threshold$ and $\altAppAbbrev \leq
\kappa$, and we claim that
\begin{align}
\label{EqnUtilBounds}  
\UtilGain{j} \leq \DelBar_j \qquad \mbox{for $j \in \{0,1 \}$.}
\end{align}
\end{subequations}
Given these inequalities and equation~\eqref{EqnCleanRestate}, in
order to prove the corollary, it suffices to show that the function
$\Phi$ is increasing in each of its four arguments.

\paragraph{Proof of the bounds\texorpdfstring{~\eqref{EqnUtilBounds}}{ (39b)}:}  We begin
by proving the upper bounds~\eqref{EqnUtilBounds}.  By the definitions
of $\UtilGain{j}$ and $\DelBar_j$, it is equivalent to prove
\begin{align*}
  \Exs [\util(\Wealth_0 + \Reward - \cost) \mid \theta \in \Theta_j]
    \leq \util(\Wealth_0 + \constR_j - \cost),
\end{align*}
where $\constR_j = \sup_{\theta \in \Theta_j} \Exs_\theta[\rfunc]$.
Now we have
\begin{align*}
  \Exs [\util(\Wealth_0 + \Reward - \cost) \mid \theta \in \Theta_j] &
  \stackrel{(i)}{\leq} \util \Big ( \Wealth_0 + \Exs[\Reward \mid
    \theta \in \Theta_j]- \cost \Big) \\
& \stackrel{(ii)}{\leq} \util \big (\Wealth_0 + \constR_j- \cost
  \big),
\end{align*}
as claimed.  Here step (i) follows by applying Jensen's inequality to
the concave function $\util$; and step (ii) follows since
$\Exs[\Reward \mid \theta \in \Theta_j] \leq \constR_j$, and the
function $\util$ is non-decreasing.

\paragraph{Proof of monotonicity:}  In order to show that
$\Phi$ is increasing of each of its four arguments, it suffices to
verify that its partial derivatives are all non-negative.  Using
$\partial_{a}$ as a shorthand for the partial derivative with respect
to a given entry $a$, we have
\begin{align*}
\partial_{\nullAppAbbrev} \Phi(\theta) & = \frac{1}{\nullAppAbbrev} +
\frac{\utilLoss + \altAppAbbrev \UtilGain{0}}{\big[ (\altAppAbbrev - \nullAppAbbrev)
    \utilLoss + \nullAppAbbrev \altAppAbbrev (\UtilGain{1} - \UtilGain{0}) \big]}, \\
\partial_{\altAppAbbrev} \Phi(\theta) & = \frac{\UtilGain{1} }{\altAppAbbrev
  \UtilGain{1} - \utilLoss} - \frac{\utilLoss + \nullAppAbbrev (\UtilGain{1}
  - \UtilGain{0})}{\big[ (\altAppAbbrev - \nullAppAbbrev) \utilLoss + \nullAppAbbrev
    \altAppAbbrev (\UtilGain{1} - \UtilGain{0}) \big]} \\
    &= \frac{\utilLoss (\utilLoss - \nullAppAbbrev \UtilGain{0})}{\big[\altAppAbbrev
  \UtilGain{1} - \utilLoss \big] \cdot \big[ (\altAppAbbrev - \nullAppAbbrev) \utilLoss + \nullAppAbbrev
    \altAppAbbrev (\UtilGain{1} - \UtilGain{0}) \big]}, \\
\partial_{\UtilGain{0}} \Phi(\theta) & = \frac{\nullAppAbbrev \altAppAbbrev}{\big[
    (\altAppAbbrev - \nullAppAbbrev) \utilLoss + \nullAppAbbrev \altAppAbbrev (\UtilGain{1} -
    \UtilGain{0}) \big]},  \\
\partial_{\UtilGain{1}} \Phi(\theta) & =  \frac{\altAppAbbrev}{\altAppAbbrev
  \UtilGain{1} - \utilLoss} - \frac{\nullAppAbbrev \altAppAbbrev}{\big[ (\altAppAbbrev - \nullAppAbbrev) \utilLoss + \nullAppAbbrev
    \altAppAbbrev (\UtilGain{1} - \UtilGain{0}) \big]} \\
    &= \frac{\altAppAbbrev^2 (\utilLoss - \nullAppAbbrev \UtilGain{0})}{\big[\altAppAbbrev
  \UtilGain{1} - \utilLoss\big] \cdot \big[ (\beta_1 - \nullAppAbbrev) \utilLoss + \nullAppAbbrev
    \beta_1 (\UtilGain{1} - \UtilGain{0}) \big]}.
\end{align*}
By definition, the terms $\nullAppAbbrev, \altAppAbbrev, \UtilGain{0}, \UtilGain{1}, \utilLoss$ are all non-negative. Furthermore, our assumptions ensure that  $\altAppAbbrev - \nullAppAbbrev > 0$ and \mbox{$\utilLoss - \nullAppAbbrev \UtilGain{0} \geq 0$}. Finally, we have previously established that \mbox{$\UtilGain{1} - \UtilGain{0} \geq 0$} and $\altAppAbbrev \UtilGain{1} - \utilLoss \geq 0$. Therefore, we conclude that all the partial derivatives are non-negative.


\subsection{Proof of\texorpdfstring{~\Cref{ThmStaircase}}{ Theorem 2}}
\label{AppProofThmStaircase} 

Recall that we have specialized to a simple-versus-simple test and
\Cref{th:bound_on_pos_null} provides a bound on the Bayes FDR in terms
of the function
\begin{align*}
\Psi(\tau) \defn \nullAppSimple \; \left \{ \frac{\altAppSimple
  \altUtilGain - \utilLoss}{(\altAppSimple - \nullAppSimple) \utilLoss
  + \nullAppSimple \altAppSimple (\altUtilGain - \nullUtilGain)}
\right \}.
\end{align*}
Given an increasing sequence of $K$ thresholds
$\{\threshold_i\}_{i=1}^K$, consider the sequence of prior null
probabilities given by
\begin{align}
  \label{EqnPriorBound}
  \priorNull^i = \frac{\beta_1(\threshold_i) \altUtilGain -
    \utilLoss}{ \beta_1(\threshold_i) \altUtilGain -
    \beta_0(\threshold_i) \nullUtilGain} \qquad \mbox{for $i = 1, \ldots, \numAgent$.}
\end{align}
By~\Cref{PropPriorElicit}, we know $\Ind(\priorNull^i; \threshold_i) =
1$ and $\fdr(\priorNull^i;\threshold_i) = \Psi(\threshold_i)$ for each
$i = 1, \ldots, K$.  Each $\priorNull^i$ is the worst-case prior null
probability with which an agent would choose to \optin for a test with
threshold $\threshold_i$, and it makes the Bayes FDR bound hold with
equality. Moreover, agents with $\priorNull^i$ will not \optin under
thresholds smaller than $\threshold_i$.

Let us first elucidate some properties of the sequence
$\{\priorNull^i\}_{i=1}^\numAgent$ from
equation~\eqref{EqnPriorBound}.  By definition, the functions
$\beta_1$ and $\beta_0$ are increasing in $\threshold$; moreover,
viewed as a function of $(\beta_0, \beta_1)$, the probabilities from
equation~\eqref{EqnPriorBound} are increasing functions in both
arguments.  (For details, see the proof
in~\Cref{AppProofPriorBtwo}). It follows that
$\{\priorNull^i\}_{i=1}^\numAgent$ is an increasing sequence,
representing agents with progressively less confidence in their
proposals.

If the opt-in population under threshold $\threshold_i$ consists
solely of agents with $\priorNull^i$, then by the sharpness guarantee
in~\Cref{ThmBound}(b), we have $\kfdr(\threshold_i) =
\fdr(\priorNull^i; \threshold_i) = \Psi(\threshold_i)$.  At the
smallest threshold $\threshold_1$, only the ``best" agents with
$\priorNull^1$ \optin, from which equality~\Bone claimed
in~\Cref{ThmStaircase} follows.

Given a mixture of agents, as we loosen the thresholds to
$\threshold_i$ for $i \geq 2$, agents with prior null probabilities
smaller than $\priorNull^i$ will also \optin. This introduces a gap
between $\kfdr(\threshold_i)$ and $\Psi(\threshold_i)$ because
\begin{align*}
  \fdr(\priorNull^j;\threshold_i) \leq \fdr(\priorNull^i;
  \threshold_i) = \Psi(\threshold_i) \qquad \mbox{for all } j < i .
\end{align*}
In order to close this gap, we construct a sequence
$\{\agentWeight{i}\}_{i=1}^K$ of mixture proportions such that the
opt-in population under $\threshold_i$ is dominated by agents with
$\priorNull^i$. Define
\begin{align*}
  \tildeagentWeight{i} \defn \frac{\agentWeight{i}
    \Ind(\priorNull^i;\threshold)}{\sum_{j=1}^K\agentWeight{j}
    \Ind(\priorNull^j;\threshold)},
\end{align*}
corresponding to the mixture proportion of agents with $\priorNull^i$
within the opt-in population under threshold $\threshold$.

We now use these weights to construct the required
sequence$\{\agentWeight{i}\}_{i=1}^K$.  We first investigate how the
ratio $\tildeagentWeight{i}$ to $\tildeagentWeight{j}$ for $j < i$
should behave so as satisfy inequality~\Btwo in~\Cref{ThmStaircase}.
We claim that it suffices if under threshold $\threshold_i$, the
opt-in proportion of ``worst-case" agents $\tildeagentWeight{i}$
satisfies
\begin{subequations}
  \begin{align}
    \label{EqnWorstCaseProp}
    \tildeagentWeight{i} \geq 1 -
    \frac{\epsilon}{\Psi(\threshold_i)}.
  \end{align}
  Once we establish this, we claim that it suffices to use an
  iterative procedure to build $\{\agentWeight{i}\}_{i=1}^K$:
  \begin{align}
    \label{EqnAbsMixtureProp}
    \agentWeight{1} = 1 \qquad \mbox{and} \qquad \agentWeight{i} =
    \bigg(\sum_{j=1}^{i-1} \agentWeight{j}\bigg) \cdot
    \frac{\Psi(\threshold_i) - \epsilon}{\epsilon} \qquad \mbox{for
      all $i \in \{2,\dots,\numAgent \}$}.
  \end{align}
\end{subequations}
The final mixture proportions $\{\agentWeight{i}\}_{i=1}^K$ are
obtained by renormalizing the weights~\eqref{EqnAbsMixtureProp} so
that they sum to one. \\

\noindent It remains to prove these two auxiliary
claims~\eqref{EqnWorstCaseProp} and~\eqref{EqnAbsMixtureProp}.

\paragraph{Proof of claim\texorpdfstring{~\eqref{EqnWorstCaseProp}}{(45a)}:}
Since $\{\priorNull^i\}_{i=1}^K$ is an increasing sequence, there are
$i$ types of agents that will \optin under threshold $\threshold_i$,
each with $\priorNull^j \leq \priorNull^i$. The gap between
$\kfdr(\threshold_i)$ and $\Psi(\threshold_i)$ is given by
\begin{align*}
  \Psi(\threshold_i) - \kfdr(\threshold_i) &= \sum_{j=1}^i
  \tildeagentWeight{j} \Psi(\threshold_i) - \sum_{j=1}^i
  \tildeagentWeight{j} \fdr(\priorNull^j;\threshold_i) \\
  & \stackrel{(i)}{=} \sum_{j=1}^{i-1} \tildeagentWeight{j}
  \big[\Psi(\threshold_i) - \fdr(\priorNull^j;\threshold_i)\big] \\
& \stackrel{(ii)}{\leq} \sum_{j=1}^{i-1} \tildeagentWeight{j}
  \Psi(\threshold_i) \\
& \stackrel{(iii)}{=} (1 - \tildeagentWeight{i}) \Psi(\threshold_i).
\end{align*}
Step (i) follows from the fact $\fdr(\priorNull^j;\threshold_j) =
\Psi(\threshold_j)$, which is our sharpness guarantee for a single
agent type. In step (ii), we use the fact that
$\fdr(\priorNull^i;\threshold_j) \geq 0$, whereas step (iii) follows
from the fact that $\sum_{j=1}^i \tildeagentWeight{j} = 1$ since
agents with $\priorNull^j > \priorNull^i$ will not \optin under
threshold $\threshold_i$.

Thus, to satisfy the inequality $\Psi(\threshold_i) -
\kfdr(\threshold_i) \leq \epsilon$, it suffices to set
\begin{align*}
\tildeagentWeight{i} \geq 1 - \frac{\epsilon}{\Psi(\threshold_i)}.
\end{align*}
We can observe that as $\threshold_i$ increases, the denominator
$\Psi(\threshold_i)$ also increases. Thus $\tildeagentWeight{i}$ must
grow progressively larger. This is because loosening the threshold
allows more agents to \optin, necessitating a higher proportion of
``worst-case" agents to offset the gap introduced by the inclusion of
``better" agents.

\paragraph{Proof of claim\texorpdfstring{~\eqref{EqnAbsMixtureProp}}{(45b)}:}
We need to construct a sequence $\{\agentWeight{i}\}_{i=1}^K$ of
mixture weights, normalized to sum to one, such that
claim~\eqref{EqnWorstCaseProp} holds. We do so by constructing a
sequence that is not normalized, and then re-normalizing.  Without
loss of generality, we can assume $\agentWeight{1} = 1$.  For
claim~\eqref{EqnWorstCaseProp} to hold, it suffices that under
$\threshold_i$, we have
\begin{align*}
\label{EqnComputeProp}
    1-\tildeagentWeight{i} &= \frac{\sum_{j=1}^{i-1}
      \agentWeight{j}}{\sum_{j=1}^{i-1} \agentWeight{j} +
      \agentWeight{i}} = \frac{\epsilon}{\Psi(\threshold_i)}.
\end{align*}
Rearranging this equation, we obtain claim~\eqref{EqnAbsMixtureProp}.


\end{document}